\documentclass[preprint,aps,showpacs]{revtex4}
\usepackage[dvips]{graphicx}
\usepackage{amsmath,amssymb}

\newcommand{\btheta} {\mbox{\boldmath$\theta$}}
\newcommand{\bphi} {\mbox{\boldmath$\phi$}}
\newcommand{\bmu} {\mbox{\boldmath$\mu$}}
\newcommand{\bnu} {\mbox{\boldmath$\nu$}}

\newcommand{\bn} {\mbox{\boldmath$n$}}
\newcommand{\bk} {\mbox{\boldmath$k$}}
\newcommand{\bp} {\mbox{\boldmath$p$}}
\newcommand{\bq} {\mbox{\boldmath$q$}}
\newcommand{\bzero} {\mbox{\boldmath$0$}}
\newcommand{\bone} {\mbox{\boldmath$1$}}
\newcommand{\btwo} {\mbox{\boldmath$2$}}
\newcommand{\bthree} {\mbox{\boldmath$3$}}
\newcommand{\bfive} {\mbox{\boldmath$5$}}
\newcommand{\bsix} {\mbox{\boldmath$6$}}

\begin{document}

\title{Moments of generalized Husimi distributions
and complexity of many-body quantum states}
\author{Ayumu Sugita\thanks{sugita@yukawa.kyoto-u.ac.jp}}
\affiliation{Yukawa Institute for Theoretical Physics (YITP), 
Kyoto University,\\
Kitashirakawa-Oiwakecho, Sakyoku, Kyoto 606-8502, Japan}
\date{\today}

\begin{abstract}
We define generalized Husimi distributions using 
generalized coherent states,
and show that their moments are good measures of
complexity of many-body quantum states.
Our construction of the coherent states is based on
the single-particle transformation group of the system.
Then the coherent states are independent-particle
states, and, at the same time, the most localized states in 
the Husimi representation.
Therefore delocalization of the Husimi distribution, which can be
measured by the moments, is a
sign of many-body correlation (entanglement).
Since the delocalization of the Husimi distribution is also related 
to chaoticity of the dynamics, it suggests
a relation between entanglement and chaos. 
Our definition of the Husimi distribution can be applied not only
to the systems of distinguishable particles, but also to those
of identical particles, i.e., fermions and bosons. 
We derive an algebraic formula to evaluate the moments of the Husimi 
distribution.
\end{abstract}

\maketitle


\section{Introduction}
Although the manifestation of chaos in quantum systems has been investigated 
extensively in the past few decades, definition of quantum 
chaos is still unclear.
Especially the studies of many-body systems are far behind those
of one-body systems. 

In \cite{sugita}, we proposed a measure of complexity 
(chaoticity) of quantum
states in one-body systems by using the Husimi distribution 
\cite{husimi}. 
The purpose of this paper is to generalize this idea to
many-body problems. We will show that the generalized Husimi
distribution is also related to quantum many-body correlation
(entanglement). Therefore it implies a relation between entanglement
and chaoticity of the dynamics.

In one-body systems, the Husimi distribution function of a quantum state 
$|\varphi\rangle $ is defined as
\footnote{
The Husimi distribution can be defined more generally as
${\cal H}(\bp, \bq) \equiv \langle \bp, \bq |\hat{\rho}|\bp, \bq\rangle$,
where $\hat{\rho}$ is the density matrix.
However, we restrict ourselves to 
pure states in this paper.}
\begin{eqnarray}
{\cal H}_{|\varphi\rangle}(\bp,\bq) \equiv 
|\langle \bp,\bq|\varphi\rangle|^{2},
\end{eqnarray}
where $|\bp,\bq\rangle$ is a coherent state 
whose average momentum and coordinate
are respectively $\bp$ and $\bq$.
Since chaoticity of classical systems can be characterized by 
delocalization of trajectories, delocalization of the Husimi
distribution can be regarded as a quantum manifestation of chaos.
We used the second moment 
as a simple measure of the delocalization. The inverse of 
the second moment represents the effective volume occupied by
the Husimi distribution, and has a good correspondence with
chaoticity of the classical system.

To extend this idea to many-body systems is not difficult, at least
formally. If we define coherent states in a many-body system,
the Husimi distribution is defined as the square of the absolute
value of the coherent state representation. 
There are several ways to generalize the idea of coherent state 
to many-body systems. 
We construct the coherent states based on the single-particle
transformation group, following the group-theoretical construction
by Perelomov \cite{perelomov}.
Then the coherent states are independent-particle states,
as we will explain below.

Let us consider a system of $m$ qubits as an example.
In this case, the single-particle (local unitary) transformation group is
$\overbrace{SU(2)\times\dots\times SU(2)}^{m}$.
Coherent states are generated by applying this group
to the ``vacuum'' \footnote{Note that the ``vacuum'' here 
may be defferent from the physical vacuum  
which is defined as the lowest energy state.  
From the group theoretical point of view, it is convenient 
to use the lowest (or the highest) weight state as the ``vacuum''. 
See section \ref{calc}.}
$|0\rangle\otimes\dots\otimes|0\rangle$.
Then we obtain all separable (disentangled) states.
Hence,
\begin{eqnarray}
\mbox{coherent state}\;
\Longleftrightarrow\;
\mbox{disentangled state}.
\label{coh-dis}
\end{eqnarray}

\begin{figure}
\includegraphics[width=0.8\textwidth]{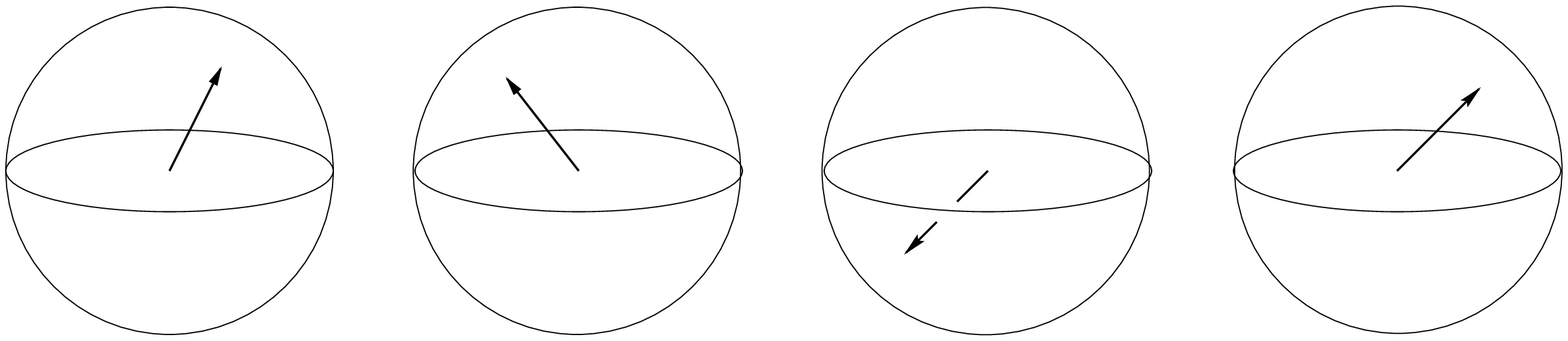}
\caption{The Husimi distribution of qubits is defined
on $S^{2}\times S^{2}\times\dots\times S^{2}$.
A disentangled state is represented by a point on this manifold,
and the Husimi distribution thereof is localized around the point.
An entangled state is represented by a delocalized distribution.}
\label{qubits_fig}
\end{figure}

The coherent states are parametrized as
\begin{eqnarray}
|\btheta, \bphi\rangle = |\theta_{1}, \phi_{1}\rangle\otimes
\dots\otimes |\theta_{m}, \phi_{m}\rangle,
\end{eqnarray}
where $|\theta, \phi\rangle$ is the Bloch sphere representation
of a qubit. The Husimi distribution of a quantum state 
$|\varphi\rangle$ is defined as
\begin{eqnarray}
{\cal H}_{|\varphi\rangle}(\btheta, \bphi) =
\left|\langle\btheta, \bphi|\varphi\rangle \right|^{2} 
\end{eqnarray}
on the ``phase space'' $S^{2}\times\dots\times S^{2}$.
A disentangled state is represented by a localized wave 
packet in the phase space. (Fig. \ref{qubits_fig})

The Husimi distribution can also be defined for systems
of identical particles, i.e., bosons and fermions.
Suppose there are $m$ identical particles
in $N$ single-particle states. 
In this case, we cannot operate 
each particle separately.
Therefore the single-particle transformation
group of this system is not 
$\overbrace{SU(N)\times\dots\times SU(N)}^{m}$, but $SU(N)$
(or $U(N)$)
\footnote{It does not matter whether 
we use $U(N)$ or $SU(N)$, since $U(N)\simeq U(1)\times SU(N)$
and the $U(1)$ subgroup corresponds to the irrelevant phase factor.
Nevertheless we prefer $U(N)$ rather than $SU(N)$
for systems of identical particles, 
because the basis of the Lie algebra of $U(N)$
is easier to write down explicitly. 
See section \ref{bosons} and \ref{fermions}.}.
The coherent states are
defined as
\begin{eqnarray}
|\zeta\rangle = U(\zeta)|0\rangle,
\end{eqnarray}
where $U(\zeta)$ is an element of $U(N)$ specified by 
the parameter $\zeta$.
The ``vacuum'' $|0\rangle$ can be written explicitly as  
\begin{eqnarray}
|0\rangle = \left\{
\begin{array}{cl}
|\varphi_{1}\rangle |\varphi_{1}\rangle \dots |\varphi_{1}\rangle
& \mbox{(boson)}\\
{\cal A}
\left(|\varphi_{1}\rangle |\varphi_{2}\rangle \dots
|\varphi_{m}\rangle\right) 
& \mbox{(fermion)}
\end{array}\right.,
\end{eqnarray}
where $|\varphi_{i}\rangle$ is the $i$-th single-particle
state and ${\cal A}$ is the anti-symmetrization operator.
Then the coherent state is
\begin{eqnarray}
|\zeta\rangle = \left\{
\begin{array}{cl}
|\varphi_{1}(\zeta)\rangle |\varphi_{1}(\zeta)\rangle \dots 
|\varphi_{1}(\zeta)\rangle
& \mbox{(boson)}\label{boson-coherent}\\
{\cal A}
\left(|\varphi_{1}(\zeta)\rangle |\varphi_{2}(\zeta)\rangle \dots
|\varphi_{m}(\zeta)\rangle\right) 
& \mbox{(fermion)}
\end{array}\right. ,
\end{eqnarray}
where $|\varphi_{i}(\zeta)\rangle = U(\zeta)|\varphi_{i}\rangle$.
In the bosonic case, the coherent states are separable,
and it is easy to see that all separable bosonic states
can be written in this form. 
Therefore
\begin{eqnarray}
\mbox{coherent state}\;
\Longleftrightarrow\;
\mbox{separable state}.
\end{eqnarray} 
In the fermionic case, a coherent state is a Slater determinant,
and any Slater determinant can be written as a coherent state. 
Namely,
\begin{eqnarray}
\mbox{coherent state}\;
\Longleftrightarrow\;
\mbox{Slater determinant}.
\end{eqnarray}
In any case, the coherent states are the least correlated ones.
We summarize the features of these
systems in Table. \ref{table1}.

The Husimi distribution is defined by using the coherent state as
\begin{eqnarray}
{\cal H}_{|\varphi\rangle}(\zeta) \equiv 
\left|\langle\zeta |\varphi\rangle\right|^{2}.
\end{eqnarray}
Since the most localized states in the Husimi representation
are the coherent states,
delocalization of the Husimi distribution implies
correlation among the particles.
The delocalization can be
measured by the R\'{e}nyi-Wehrl entropy \cite{zyczkowski}
\begin{eqnarray}
S^{(q)}_{|\varphi\rangle} \equiv \frac{1}{1-q}\ln M^{(q)}_{|\varphi\rangle},
\end{eqnarray}
where $M^{(q)}$ is the moment with an index $q>0$:
\begin{eqnarray}
M^{(q)}_{|\varphi\rangle} =
 \int d\mu (\zeta ) 
\left\{{\cal H}_{|\varphi\rangle}(\zeta )\right\}^{q}.
\label{moment}
\end{eqnarray}
Here, $d\mu (\zeta)$ is the Haar measure of the group.
Hence $M^{(q)}$ is an invariant of the group transformation. 
Note that $S^{(q)}$ reproduces the normal entropy in the limit
$q\rightarrow 1$:
\begin{eqnarray}
\lim_{q\to 1}S^{(q)}_{|\varphi\rangle} = 
- \int\! d\mu(\xi)\; 
{\cal H}_{|\varphi\rangle}(\zeta) \ln {\cal H}_{|\varphi\rangle}(\zeta).
\end{eqnarray}

The R\'{e}nyi-Wehrl entropy represents the effective volume
occupied by the Husimi distribution. For instance, if the Husimi 
distribution takes the same value over a region with volume $V$
and takes zero value outside of it
\footnote{In fact, no state gives such a distribution. This is an
example only to explain the meaning of the R\'{e}nyi-Wehrl entropy.}
, $S^{(q)}=\ln V$.
Therefore it is natural to expect that $S^{(q)}$ takes the minimum value
for the coherent states. 
This conjecture (generalized Lieb-Wehrl conjecture)
was formed in \cite{zyczkowski}, following the questions
raised by Wherl \cite{wehrl} and Lieb \cite{lieb}.
For integer indices $q\ge 2$ it was proved in \cite{sugita2}.
According to the conjecture,
\begin{eqnarray}
S^{(q)}_{|\varphi\rangle} = S^{(q)}_{min}\;
\Longleftrightarrow\;
|\varphi\rangle:\, 
\left\{\begin{array}{cc}
\mbox{separable} & \mbox{(qubits, bosons)}\\
\mbox{Slater determinant} & \mbox{(fermions)}
\end{array}\right. .
\end{eqnarray}
Thus $S^{(q)}$ shows whether the particles are correlated or not.
If $S^{(q)}$ is large, the Husimi distribution is delocalized,
and we need a lot of wave packets (independent-particle states) 
to represent the state. Threfore the R\'{e}nyi-Wehrl entropy 
can be regarded as a measure of complexity.

Note that the complexity we measure here is completely independent
of the complexity of single-particle states which was the subject
of \cite{sugita}. For instance, we can construct a Slater determinant
from highly chaotic single-particle states, which is still the most
localized state in the many-body Husimi distribution.

When a Hamiltonian of the system is specified, the time evolution
operator of the system can be represented by the coherent state
path integral. The stationary phase condition of this integral
leads to an equation of motion in the phase space, which defines
the ``classical'' dynamics of the system. (See, for example, \cite{zhang}.) 
In the case of qubits, for example, 
the equation of motion is that of classical spins.
When the ``classical'' motion is chaotic, the Husimi distribution
will probably spread all over the phase space. 
This means that chaotic dynamics leads to highly complex
states. 

Next we consider how to calculate the moments (\ref{moment}).
Although the definition of the moments (\ref{moment}) 
using the Haar measure is an elegant way to obtain
invariants of the group, this integral
representation is almost useless in numerical calculations 
because the dimension of the phase space is huge when
there are many degrees of freedom.
Since the coherent states form an overcomplete set,
the coherent state representation is highly redundant.
We should avoid such redundancy in real calculations,
as we did in \cite{sugita} for one-body problems. 

A complex quantum state is usually represented by expansion
coefficients in a basis. Therefore it is desirable to represent
the moments (\ref{moment}) directly by them.
Fortunately, it is possible if the index $q$ is a positive integer. 
We will derive an algebraic formula which represents the moment
directly by the expansion coefficients 
in section \ref{calc}. This explicit formula is not only practically
useful, but also theoretically insightful.

\begin{table*}
\begin{tabular}{ccccc}
\hline\hline
 & one-body system & bosons & fermions & distinguishable particles \\
 & (n dim.)        & $B[N,m]$ & $F[N,m]$   & $D[N_{1},\dots,N_{m}]$\\
\hline
group & HW group  & $U(N)$
& $U(N)$ 
& $SU(N_{1})\times\dots\times SU(N_{m})$\\ 
\hline
phase space & $\mathbb{R}^{2n}$ & $\mathbb{C}P^{N-1}$ & 
$G_{m,N}(\mathbb{C})$ & 
$\mathbb{C}P^{N_{1}-1}\times\dots\times \mathbb{C}P^{N_{m}-1}$ \\
\hline
coherent state 
& Gaussian & separable state & Slater Det. & separable state \\
& wave packet & $|\varphi\rangle|\varphi\rangle\dots|\varphi\rangle$
&${\cal A}(|\varphi_{1}\rangle|\varphi_{2}\rangle\dots|\varphi_{m}\rangle)$&
$|\varphi_{1}\rangle|\varphi_{2}\rangle\dots|\varphi_{m}\rangle$\\
\hline\hline
\end{tabular}
\caption{Coherent states for one-body systems, bosons, fermions and 
distinguishable particles.
$B[N,m]$ ($F[N,m]$) means a system of 
$m$ bosons (fermions) in $N$ single-particle states. 
$D[N_{1},\dots,N_{m}]$ is a system of $m$ distinguishable
particles in which $i$-th particles takes $N_{i}$ states.
$\mathbb{C}P^{n}$ is the $n$-dimensional complex projective space 
and $G_{m,N}(\mathbb{C})\simeq U(N)/\left(U(N-m)\times U(m)\right)$ 
is the Grassmann manifold. Note that $\mathbb{C}P^{1}\simeq S^{2}$.
HW group in the second row means the Heisenberg-Weyl group,
which is generated by $\hat{\bp}$ and $\hat{\bq}$.} 
\label{table1}
\end{table*}

This paper is organized as follows. In section \ref{calc}
we derive the explicit formula for the moments.
In this formula, the moments are represented by 
the expansion coefficients of the state and group-theoretical factors
which are independent of the state.
The main concern in the following
sections is to determine the group-theoretical factors.
After explaining the general formalism, we consider 
bosonic systems with two single-particle
states in detail in section \ref{twoboson}. 
These systems have $U(2) \simeq U(1)\times SU(2)$ as
the single-particle transformation group, 
and the corresponding phase space is the two-demensionasl sphere 
$S^{2}$. Therefore we can visualize the Husimi distribution in
this case, which  
will help us understand the idea of the Husimi distribution in
many-body systems.
Based on this analysis, we treat
bosonic systems generally in section \ref{bosons}. 
In section \ref{fermions} we treat fermionic systems. 
In section \ref{distinguishable}, we consider distinguishable particles
including qubits. We investigate the second moment in
2- and 3-qubit cases in detail, and show the relation
between the second moment and other known measures of entanglement.
The final section is devoted to concluding remarks.

\section{Explicit formula for the moments
with integer indices}
\label{calc}

In this section, we show how to obtain an explicit formula
for the moment $M^{(q)}$ for a positive integer $q$. 
Our method is based on the expression 
of the moments (\ref{projection}) using the tensor products of the state,
which was used also in \cite{sugita2} to prove the generalized
Lieb-Wehrl conjecture. We briefly review the derivation of 
(\ref{projection}), and show how the formula is obtained from it.

To begin with, we briefly state the definition of the generalized 
coherent states 
of a compact semisimple Lie group $G$ \cite{perelomov}.
The Lie algebra of $G$ can be written in Cartan basis
$\{E_{\alpha}, H_{j}\}$. Let $D_{\lambda}$
be an irreducible representation space of $G$ with the
lowest weight $-\lambda$. Coherent states in $D_{\lambda}$ are obtained by the action of the
group $G$ on the lowest weight state $|-\lambda\rangle$,
which can be written explicitly as
\begin{eqnarray}
|\zeta, \lambda\rangle \equiv {\cal N}(\zeta)
\exp (\zeta_{\alpha}E_{\alpha})|-\lambda\rangle.
\label{def_coh}
\end{eqnarray}
Here, ${\cal N}(\zeta)$ is the normalization constant
and $\alpha$ runs over all positive roots. 

Next we consider the moment with a positive integer index $q$
\begin{eqnarray}
M^{(q)}_{|\varphi\rangle} = c_{\lambda}\int d\mu (\zeta)\, 
|\langle \zeta, \lambda|\varphi\rangle |^{q}.
\end{eqnarray}
Here, $d\mu (\zeta)$ is the Haar measure of $G$ and
$c_{\lambda}={\rm dim}D_{\lambda}$ is the normalization constant.
The moment can be rewritten as
\begin{eqnarray}
M^{(q)}_{|\varphi\rangle} = c_{\lambda}\int d\mu(\zeta)\, 
\left(\langle\varphi |^{\otimes q}\right)
\left(|\zeta, \lambda\rangle^{\otimes q}\right)
\left(\langle \zeta, \lambda|^{\otimes q}\right)
\left(\varphi\rangle^{\otimes q}\right).
\label{moment_tensor}
\end{eqnarray}

The important point to note is that the tensor product of
the coherent state $|\zeta\rangle^{\otimes q}$ is again a
coherent in another irreducible representation 
$D_{q\lambda}\subset D_{\lambda}^{(q)}$
whose lowest weight is $-q\lambda$.
It can be shown explicitly as
\begin{eqnarray}
|\zeta, \lambda\rangle^{\otimes q} = 
{\cal N}(\zeta)^{q}
\exp \left(\zeta_{\alpha}E^{(q)}_{\alpha}\right)|-\lambda\rangle^{\otimes q}
= |\zeta, q\lambda\rangle. 
\end{eqnarray}
Here, $E^{(q)}_{\alpha}$ is the representation of $E_{\alpha}$
in $D_{\lambda}^{\otimes q}$, whose explicit form is
\begin{eqnarray}
E^{(q)}_{\alpha} = 
E_{\alpha}\otimes I\otimes\dots\otimes I +
I\otimes E_{\alpha}\otimes\dots\otimes I 
+ \dots +
I\otimes\dots\otimes I\otimes E_{\alpha}.
\end{eqnarray}
It is easy to verify that $|-\lambda\rangle^{\otimes q}$ is the
lowest weight state of $D_{q\lambda}$.

Since the ``resolution of unity'' 
\begin{eqnarray}
I = c_{q\lambda}\int d\mu (\zeta) 
|\zeta, q\lambda\rangle\langle\zeta, q\lambda| 
\end{eqnarray}
holds in $D_{q\lambda}$, (\ref{moment_tensor}) can be written as
\begin{eqnarray}
M_{|\varphi\rangle}^{(q)} =
\frac{{\rm dim}\, D_{\lambda}}{{\rm dim}\, D_{q\lambda}}
\left(\langle\varphi|^{\otimes q}\right)P_{D_{q\lambda}}
\left(|\varphi\rangle^{\otimes q}\right),
\label{projection}
\end{eqnarray}
where $P_{D_{q\lambda}}$
is the projection operator to $D_{q\lambda}$.
For our purpose, it is convenient to introduce normalized
moments
\begin{eqnarray}
\tilde{M}^{(q)}_{|\varphi\rangle} 
\equiv 
\frac{{\rm dim}\, D_{q\lambda}}{{\rm dim}\, D_{\lambda}}\,
M_{|\varphi\rangle}^{(q)}
= \langle P_{D_{q\lambda}}\rangle,
\end{eqnarray}
where the symbol $\langle\rangle$ denotes the 
expectation value for $|\varphi\rangle^{\otimes q}$.
Then $M^{(q)}_{|\varphi\rangle}=1$ if $|\varphi\rangle$
is a coherent state, and 
$M^{(q)}_{|\varphi\rangle}<1$ if it is not.

If the state is expressed as
\begin{eqnarray}
|\varphi\rangle = \sum_{i}c_{i}|i\rangle
\end{eqnarray}
with an orthonormal basis $\{|i\rangle\}$ in $D_{\lambda}$,
the normalized $q$-th moment is
\begin{eqnarray}
\tilde{M}_{|\varphi\rangle}^{(q)} =
\sum_{j}|B_{j}|^{2},
\label{general1}
\end{eqnarray} 
\begin{eqnarray}
B_{j} = \sum_{i_{1},\dots,i_{q}}
\langle j, q\lambda|i_{1},\dots,i_{q}\rangle 
c_{i_{1}}\dots c_{i_{q}},
\label{general2}
\end{eqnarray}
where $\{|j, q\lambda\rangle\}$ is an orthonormal basis
of $D_{q\lambda}$, and 
\begin{eqnarray}
\langle j, q\lambda |i_{1},\dots,i_{q}\rangle 
\equiv
\langle j, q\lambda |
\left(|i_{1}\rangle\otimes |i_{2}\rangle \otimes \dots\otimes
 |i_{q}\rangle\right).
\end{eqnarray}
Thus we have obtained the explicit formula for the moments.

Since the terms in the sum (\ref{general2}) are invariant 
under permutations of $i_{1},i_{2},\dots,i_{q}$,
the same term appears $q!$ times if all indices are different.
In general, the same indices appears in the set $\{i_{s}\}$.
We represent it as
as $(i_{1}^{k_{1}},i_{2}^{k_{2}},\dots,i_{l}^{k_{l}})$, which 
means the index $i_{l}$ appears $k_{l}$ times. Then 
this term appears $\binom{q}{k_{1},k_{2},\dots,k_{l}}$ times in the 
sum (\ref{general2}), where 
\begin{eqnarray}
\binom{q}{k_{1},k_{2},\dots,k_{l}} \equiv \left\{
\begin{array}{cc}
\frac{q!}{k_{1}!\,k_{2}!\,\dots\,k_{l}!} & 
(\mbox{if}\,k_{1}+k_{2}+\dots+k_{l}=q)\\
0 & (\mbox{else})
\end{array}\right.
\end{eqnarray}
is the multinomial coefficient.
Therefore (\ref{general2}) can be rewritten as
\begin{eqnarray}
B_{j} = \sum \binom{q}{k_{1},k_{2},\dots,k_{l}}
\langle j,q\lambda |i_{1}^{k_{1}},i_{2}^{k_{2}},
\dots,i_{l}^{k_{l}}\rangle c_{i_{1}}^{k_{1}}c_{i_{2}}^{k_{2}}\dots
c_{i_{l}}^{k_{l}},
\label{general3}
\end{eqnarray}
where 
\begin{eqnarray}
\langle j,q\lambda |i_{1}^{k_{1}},i_{2}^{k_{2}},
\dots,i_{l}^{k_{l}}\rangle \equiv \langle j,q\lambda |
\left(|i_{1}\rangle^{\otimes k_{1}}\otimes 
|i_{2}\rangle^{\otimes k_{2}} \otimes \dots\otimes
 |i_{q}\rangle^{\otimes k_{l}}\right).
\end{eqnarray}
The sum in (\ref{general3}) is taken over all possible combinations of
$\{i_{s}\}$ and $\{k_{s}\}$.

Let the irreducible decomposition of $D_{\lambda}^{\otimes q}$ be 
\begin{eqnarray}
D_{\lambda}^{\otimes q} = \sum_{\nu,\tau}
D_{\nu,\tau_{\nu}}.
\end{eqnarray}
Here, $D_{\nu, \tau_{\nu}}$ is an irreducible representation with
the lowest weight $-\nu$, and the additional integer suffix
$\tau_{\nu}$ is the multiplicity label.
Then
\begin{eqnarray}
\sum_{\nu,\tau_{\nu}}\langle P_{\nu,\tau_{\nu}}\rangle = {\cal N}^{2q}
\end{eqnarray} 
holds, where $P_{\nu,\tau_{\nu}}$ is the projection operator
to $D_{\nu, \tau_{\nu}}$ and
${\cal N}=\left(\sum_{i}|c_{i}|^{2}\right)^{1/2}$ is the norm of
$|\varphi\rangle$ which is usually normalized to $1$.
Hence
\begin{eqnarray}
\tilde{M}^{(q)}_{|\varphi\rangle} = 
1 - \sum_{\nu\ne q\lambda, \tau_{\lambda}} \langle
P_{\nu,\tau_{\nu}}\rangle.
\label{decompose}
\end{eqnarray}
Note that $D_{q\lambda}$ is multiplicity-free.
In many cases, this formula 
is more useful and
informative than calculating $\langle P_{q\lambda}\rangle$
directly, as we will see later.


\section{Simple example: bosons in two single-particle states}
\label{twoboson}

\subsection{Algebra of generators}
Let us consider a system of $m$ bosons in two single-particle
states as a simple example.
In this case, there are $m+1$ many-body states, which
are  
$|0,m\rangle, |1,m-1\rangle,...,|m,0\rangle$ in the occupation
number representation. 
The many-body states are generated by applying
particle-hole excitation operators 
\begin{eqnarray}
X_{i}^{j} = a_{i}a_{j}^{\dagger} \;\;\; (1\le i,j,\le 2)
\end{eqnarray}
to the ``vacuum'' $|0,m\rangle$.
Since the total number operator 
$\hat{N}=a_{1}^{\dagger}a_{1}+a_{2}^{\dagger}a_{2}$
corresponds to the irrelevant total phase, we can
remove it from the algebra.
The rest forms the Lie algebra of $SU(2)$:
\begin{eqnarray}
J_{+} & = & a_{1}a_{2}^{\dagger}, \\
J_{-} & = & a_{2}a_{1}^{\dagger}, \\
J_{z} & = & \frac{1}{2}
(a_{2}a_{2}^{\dagger} - a_{1}a_{1}^{\dagger}). 
\label{su2}
\end{eqnarray} 
Actually, it is easy to verify the following
commutation relations
\begin{eqnarray}
[J_{z},J_{\pm}] = \pm J_{\pm},\;\;\; [J_{+},J_{-}] = 2J_{z}.
\end{eqnarray}

The two single-particle states $|0,1\rangle$ and $|1,0\rangle$ form
the fundamental (spin $1/2$) representation of $SU(2)$.
The $m$-particle states form an irreducible representation
with spin $m/2$, which we denote as $D_{m/2}$. 
Hereafter, we mainly use spin quantum numbers $j$ and $j_{z}$,
instead of occupation numbers,
to specify a quantum
state. We put suffixes $n$ and $s$
to distinguish the two notations. The relation between the two 
notations is
\begin{eqnarray}
|n_{1} ,n_{2}\rangle_{n} = 
\left|j=\frac{n_{1}+n_{2}}{2}, j_{z}=\frac{n_{1}-n_{2}}{2}\right\rangle_{s}.
\end{eqnarray}

\subsection{Coherent state and the Husimi distribution}
The $SU(2)$ coherent states in the spin $j$ representation are 
defined as \cite{perelomov}
\begin{eqnarray}
|j,\zeta\rangle = {\cal N}(\zeta)\exp (\zeta J_{+})|j,-j\rangle_{s},
\label{def}
\end{eqnarray}
where ${\cal N}(\zeta)$ is a normalization factor.
In our case, the total spin $j$ is determined by the number of
particles by $j=m/2$.

$\zeta$ is considered to be a coordinate system 
of $\mathbb{C}P^{1}\simeq S^{2}$,
which is related to the angular variables of the sphere
$(\theta,\varphi)$ by
\begin{eqnarray}
\zeta & = & -e^{-i\phi}\tan\frac{\theta}{2}.
\end{eqnarray}
The coherent state $|j,\zeta\rangle$ is expanded as
\begin{eqnarray}
|j,\zeta\rangle & = & \sum_{\mu=-j}^{j}u_{j,\mu}(\zeta)
|j, \mu \rangle_{s},
\label{su2coherent}
\end{eqnarray}
where
\begin{eqnarray}
u_{j,\mu}(\zeta)
&  = &
\binom{2j}{j+\mu}^{1/2}
e^{-i(j+\mu)\phi}
\left(-\sin\frac{\theta}{2}\right)^{j+\mu}
\left(\cos\frac{\theta}{2}\right)^{j-\mu}.
\end{eqnarray}

\begin{figure}
\includegraphics[height=0.35\textwidth]{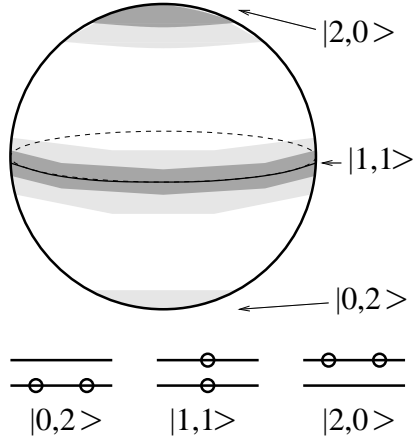}
\begin{center}
\end{center}
\caption{Rough sketch of the Husimi distribution for bosons 
with $N=m=2$.
$|2,0\rangle_{n} $ and $|0,2\rangle_{n}$ are coherent states, but
$|1,1\rangle_{n}$ is not. Therefore $|1,1\rangle_{n}$ has a rather
extended distribution along the equator, whose normalized second moment is
the minimum value $2/3$. This result shows that 
$|1,1\rangle_{n}$ can not be written as a tensor 
product of single-particle states.}
\label{sphere}
\end{figure}

The Husimi distribution function of a $m$-particle state $|\varphi\rangle$ 
is defined as 
\begin{eqnarray}
{\cal H}_{|\varphi\rangle}(\zeta) = 
|\langle j=m/2,\zeta |\varphi\rangle|^{2}.
\end{eqnarray}
For example, let us consider the simplest nontrivial case
$m=2$ which corresponds to the spin 1 representation.
The Husimi distributions of the three basis states are
\begin{eqnarray}
{\cal H}_{|2,0\rangle_{n}}(\theta,\phi) 
& = & \sin^{4}\frac{\theta}{2},\\
{\cal H}_{|1,1\rangle_{n}}(\theta,\phi)  & = & 
2\sin^{2}\frac{\theta}{2}\cos^{2}\frac{\theta}{2}
= \frac{1}{2}\sin^{2}\theta,\\
{\cal H}_{|0,2\rangle_{n}}(\theta,\phi)  & = & \cos^{4}\frac{\theta}{2},
\end{eqnarray}
which are localized around the north pole ($\theta=\pi$),
the equator ($\theta = \pi/2$) and the south pole ($\theta=0$)
respectively. (See Fig. \ref{sphere}.)
Among the three states, $|2,0\rangle_{n}$ and $|0,2\rangle_{n}$
are separable, but
\begin{eqnarray}
|1,1\rangle_{n} = \frac{1}{\sqrt{2}}
(|1,0\rangle_{n}|0,1\rangle_{n} + |0,1\rangle_{n}|1,0\rangle_{n})
\end{eqnarray}
is not. 
Coresponding to this fact, the Husimi distribution of
$|1,1\rangle_{n}$ is broader than 
the other two.

\subsection{moments}
The $q$-th moment of this distribution is defined as
\begin{eqnarray}
M^{(q)}_{|\varphi\rangle} 
= (2j+1)\int d\mu(\zeta)\; 
\left\{{\cal H}_{|\varphi\rangle}(\zeta)\right\}^{q},
\label{m2_su2}
\end{eqnarray}
where $d\mu(\xi)$ is the Haar measure 
\begin{eqnarray}
d\mu(\xi) = \frac{1}{4\pi}
\sin\theta d\theta d\phi .
\end{eqnarray}
$M^{(q)}_{|\varphi\rangle}$ is invariant under 
the $SU(2)$ transformation by definition of the Haar measure.

We can calculate the moments according to the general formula 
(\ref{general1}) and (\ref{general2}). 
Here we consider the second moment, which can be calculated from
the CG (Clebsch-Gordan) series
\begin{eqnarray}
D_{m/2}\otimes D_{m/2} = D_{m}\oplus D_{m-1}\oplus\dots\oplus D_{0}.
\end{eqnarray} 
If the state $|\varphi\rangle$ is represented as
\begin{eqnarray}
|\varphi\rangle =
\sum_{\nu=-m/2}^{m2/}c_{\nu}|m/2,\nu\rangle_{s},
\end{eqnarray}
the normalized second moment is
\begin{eqnarray}
\tilde{M}^{(2)} = \frac{2m+1}{m+1}M^{(2)} =
\sum_{\mu=-m}^{m} |B_{\mu}|^{2},
\end{eqnarray}
\begin{eqnarray}
B_{\mu} = \langle m,\mu| m/2,m/2; \nu,\nu'\rangle c_{\nu}c_{\nu'}.
\end{eqnarray}
Here, $\langle m,\mu| m/2,m/2; \nu,\nu'\rangle$ is the CG
coefficient, whose explicit form is
\begin{eqnarray}
\langle 2j,\mu|j,j; \nu,\nu'\rangle =
\sqrt{\frac{\binom{2j}{j+\nu}\binom{2j}{j+\nu'}}
{\binom{4j}{2j+\mu}}}\delta_{\mu,\nu+\nu'}.
\end{eqnarray}

Let us examine some simple cases. If $m=1$ ($j=1/2$), 
\begin{eqnarray}
\tilde{M}^{(2)}_{|\varphi\rangle} = \langle P_{1}\rangle
= \left(|c_{1/2}|^{2}+|c_{-1/2}|^{2}\right)^{2}=1,
\end{eqnarray}
where $P_{j}$ denotes the projection
operator to the spin $j$ representation.
This result is natural because this is a 
one-body system and hence there is no many-body correlation.
In the simplest non-trivial case $m=2$ ($j=1$), the normalized 
second moment is
\begin{eqnarray}
\tilde{M}^{(2)} = \langle P_{2}\rangle = 
|c_{1}|^{4} + |c_{-1}|^{4}
+2|c_{1}c_{0}|^{2}+2|c_{0}c_{-1}|^{2}
+ \frac{2}{3}|c_{1}c_{-1} + c_{0}^{2}|^{2}.
\label{p2}
\end{eqnarray}
We can also write it as
\begin{eqnarray}
\tilde{M}^{(2)} = 1 - \langle P_{1}\rangle - \langle P_{0}\rangle
\end{eqnarray}
according to (\ref{decompose}). $\langle P_{1}\rangle$
vanishes identically because the spin 1 part is anti-symmetric
with respect to the exchange of the two components of the
tensor product. The spin 0 part is
\begin{eqnarray}
|0,0\rangle_{s} = \frac{1}{\sqrt{3}}
\left(|1,1\rangle_{s}|1,-1\rangle_{s} 
- |1,0\rangle_{s}|1,0\rangle_{s}
+ |1,-1\rangle_{s}|1,1\rangle_{s}\right).
\end{eqnarray}
Hence
\begin{eqnarray}
\tilde{M}^{(2)}_{|\varphi\rangle} = 
1 - \frac{1}{3}|c_{0}^{2}-2c_{1}c_{-1}|^{2}.
\end{eqnarray}

The maximum value of $\tilde{M}^{(2)}$ is $1$, which is obtained when
$c_{0}^{2}=2c_{1}c_{-1}$. This is the condition for the 
coherent states.
The minimum value is $2/3$, which is obtained if and only if
\begin{eqnarray}
\Im \left(c_{0}^{2}c_{1}^{*}c_{-1}^{*}\right) & = & 0,\\
\Re \left(c_{0}^{2}c_{1}^{*}c_{-1}^{*}\right) & \le & 0,
\end{eqnarray}
\begin{eqnarray}
|c_{1}| & = & |c_{-1}|.
\end{eqnarray}
For example, $|1,1\rangle_{n}$ satisfies this condition.

\section{General treatment of bosonic systems}
\label{bosons}
In this section we treat bosonic systems generally.

\subsection{Coherent states for bosonic systems}
Suppose there are $m$ bosons in $N$ single-particle states.
The particle-hole excitation operators
\begin{eqnarray}
X_{i}^{j} = a_{i}a_{j}^{\dagger}\;\;\; (1\le i,j\le N)
\end{eqnarray}
satisfy the commutation relations of $\mathfrak{u}(N)$:
\begin{eqnarray}
[X_{i}^{j}, X_{l}^{k}] = 
X_{l}^{j}\delta_{i}^{k} -
X_{i}^{k}\delta_{l}^{j}.
\label{commut}
\end{eqnarray}
These operators generates the single-particle
transformation:
\begin{eqnarray}
[X_{i}^{j}, a_{k}] = \delta_{k}^{j}a_{i}.
\label{generate}\end{eqnarray}

\begin{figure}
\begin{picture}(220,90)(0,0)
\thicklines
\put(0,70){\framebox(15,15){}}
\put(15,70){\framebox(15,15){}}
\put(30,70){\framebox(30,15){$\cdot\cdot\cdot$}}
\put(60,70){\framebox(15,15){}}
\put(110,70){\framebox(15,15){}}
\put(125,70){\framebox(15,15){}}
\put(140,70){\framebox(90,15){$\cdot\cdot\cdot\cdot\cdot$}}
\put(230,70){\framebox(15,15){}}
\put(35,40){\makebox(0,0){$[1^{m}]=[\overbrace{1,1,...,1}^{m}]$}}
\put(175,40){\makebox(0,0){$[1^{qm}]=[\overbrace{1,1,\dots\dots,1}^{qm}]$}}
\put(35,10){\makebox(0,0){$B[N,m]$}}
\put(175,10){\makebox(0,0){$B^{(q)}[N,m] = B[N,qm]$}}
\end{picture}
\caption{Irreducible representations of $U(N)$,
$B[N,m]=B^{(1)}[N,m]$ and $B^{(q)}[N,m]$. }
\label{young_boson}
\end{figure}
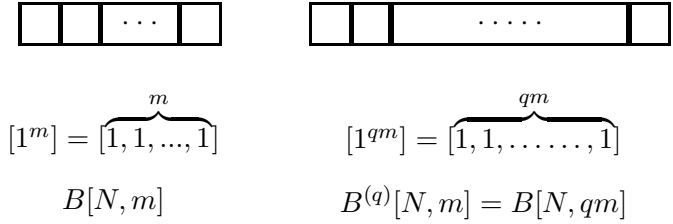

The $N$ single-particle states
form the fundamental representation of $U(N)$, and
the bosonic many-body states form an irreducible representation 
which consists of symmetric combinations of the single-particle states. 
In the Young diagram, 
this representation is $[1^{m}]$ 
(See Fig. \ref{young_boson}), 
and we denote it as $B[N,m]$. We will
drop $[N,m]$ when it is obvious.

A coherent state is obtained by
applying the $U(N)$ transformation to the bosonic
``vacuum'' $|0\rangle = |0,\dots,0,m\rangle$. 
As we have explained in the introduction, the bosonic
coherent states are of the form
\begin{eqnarray}
|\zeta\rangle = |\varphi(\zeta)\rangle|\varphi(\zeta)\rangle\dots
|\varphi(\zeta)\rangle
\label{b_coh}
\end{eqnarray}
where $|\varphi(\zeta)\rangle = U(\zeta)|\varphi_{1}\rangle$.
Unitary transformations in the subspace
spanned by $|\varphi_{2}\rangle, \dots ,|\varphi_{N}\rangle$
is irrelevant for the definition of the coherent state (\ref{b_coh}),
and the phase transformation of $|\varphi_{1}\rangle$ is also irrelevant.
Therefore the coherent states are specified by a point on
the manifold
\begin{eqnarray}
U(N)/\left(U(1)\times U(N-1)\right) \simeq \mathbb{C}P^{N-1}.
\end{eqnarray} 
An explicit form of the parametrization of the coherent states is
\begin{eqnarray}
|\zeta\rangle = {\cal N}(\zeta) 
\exp \left(\sum_{j=2}^{N}\zeta_{j}X_{1}^{j}\right)|0\rangle,
\end{eqnarray}
where $\zeta = (\zeta_{2},\zeta_{3},\dots,\zeta_{N})$ is a coordinate 
system of $\mathbb{C}P^{N-1}$.

The stationary phase condition of the coherent state path integral
leads to an equation of motion on the phase space $\mathbb{C}P^{N-1}$,
which describes the dynamics of the
condensed bosons.

\subsection{Moments of the Husimi distribution}
The Husimi distribution is defined as
\begin{eqnarray}
{\cal H}_{|\varphi\rangle}(\zeta) = 
|\langle \zeta|\varphi\rangle|^{2}.
\end{eqnarray}
The normalized $q$-th moment thereof is 
\begin{eqnarray}
\tilde{M}^{(q)}_{|\varphi\rangle} = 
\frac{{\rm dim}\, D_{q\lambda}}{{\rm dim}\, D_{\lambda}}\,
M_{|\varphi\rangle}^{(q)}
= 
({\rm dim}\, D_{q\lambda})\int d\mu (\zeta)\,
 \left\{{\cal H}_{|\varphi\rangle}(\zeta)\right\}^{q}. 
\end{eqnarray} 
Here, $D_{\lambda}$ is the bosonic Hilbert space $B[N,m] = [1^{m}]$,
and  $D_{q\lambda}$ corresponds to $[1^{qm}]$, which we denote
as $B^{(q)}[N,m]$. (See Fig. \ref{young_boson}.)
Its dimension is 
\begin{eqnarray}
{\rm dim} B^{(q)}[N,m] = \frac{(N+qm-1)!}{(N-1)!\,(qm)!},
\end{eqnarray}
hence
\begin{eqnarray}
\frac{{\rm dim}\, D_{q\lambda}}{{\rm dim}\, D_{\lambda}}
=
\frac{\Gamma(N+qm)\Gamma (m+1)}{\Gamma (N+m)\Gamma (qm+1)}.
\end{eqnarray}

We can calculate the normalized moments by using the general formulae
(\ref{general1}) and (\ref{general2}).
Let us consider the simplest case $q=2$. The basis states of $B^{(2)}$
can be constructed as
\begin{eqnarray}
|B^{(2)},\bk\rangle & = & 
C\prod_{j=2}^{N}\left\{{X^{(2)}}_{1}^{j}\right\}^{k_{j}}
|0\rangle_{m}\otimes |0\rangle_{m}.
\label{B2}
\end{eqnarray}
Here, $\bk = (k_{1},k_{2},\dots,k_{N})$ represents the occupation
numbers, 
where the total number $|\bk|\equiv \sum_{j}k_{j}$ is
constrained to $2m$.
$C$ is a normalization constant, $|0\rangle_{m}
\equiv|0,\dots,0,m\rangle$ is the $m$-particle vacuum, and 
\begin{eqnarray}
{X^{(2)}}_{i}^{j}\equiv X_{i}^{j}\otimes I + I\otimes X_{i}^{j}
\end{eqnarray}
is a generator of the Lie algebra in $B^{\otimes 2}$.
By expanding (\ref{B2}), we obtain
\begin{eqnarray}
|B^{(2)},\bk\rangle & = & 
C\prod_{j=2}^{N}
\left\{\sum_{n_{j}=0}^{k_{j}}\binom{k_{j}}{n_{j}}
(X_{1}^{j})^{n_{j}}\otimes(X_{1}^{j})^{k_{j}-n_{j}}\right\}
|0\rangle_{m}\otimes |0\rangle_{m}\\
& = &
C\sum_{n_{2}=0}^{k_{1}}\dots \sum_{n_{N}=0}^{k_{N}}
\left\{\prod_{j=2}^{N}\binom{k_{j}}{n_{j}}
(X_{1}^{j})^{n_{j}}\otimes (X_{1}^{j})^{k_{j}-n_{j}}\right\}
|0\rangle_{m}\otimes|0\rangle_{m},
\label{before}
\end{eqnarray}
where
\begin{eqnarray}
\prod_{j=2}^{N}(X_{1}^{j})^{n_{j}}|0\rangle_{m} & = &
\left(a_{N}^{\dagger}\right)^{n_{N}}
\left(a_{N-1}^{\dagger}\right)^{n_{N-1}}\dots
\left(a_{2}^{\dagger}\right)^{n_{2}}
a_{1}^{n_{N}+\dots +n_{2}}
\frac{\left(a_{1}^{\dagger}\right)^{m}}{\sqrt{m!}}|0\rangle_{0}\\
& = & 
\frac{1}{\sqrt{m!}}\frac{m!}{n_{1}!}\prod_{j=1}^{N}
\left(a_{j}^{\dagger}\right)^{n_{j}}|0\rangle_{0}\\
& = &
\frac{\sqrt{m!}}{n_{1}!}\sqrt{\bn !}|\bn\rangle.
\label{after}
\end{eqnarray}
Here, $\bn=(n_{1},\dots,n_{N})$ represents the occupation numbers,
where $n_{1}\equiv m-(n_{N}+\dots+n_{2})$. The factorial
of the vector $\bn$ means 
\begin{eqnarray}
\bn ! \equiv \prod_{j=1}^{N}n_{j}!.
\end{eqnarray} 
By substituting (\ref{after}) into (\ref{before}), we obtain
\begin{eqnarray}
|B^{(2)},\bk\rangle & = & 
C'\sum_{|\bn|=m}
\sqrt{\frac{\bk !}{\bn ! (\bk-\bn)!}}\,|\bn\rangle\otimes |\bk-\bn\rangle.
\end{eqnarray}
The normalization constant $C'$ is determined by
\begin{eqnarray}
\frac{1}{C'^{2}} & = & \sum_{|\bn|=m}\frac{\bk !}{\bn !(\bk-\bn)!}\\
& = &
 \frac{(2m)!}{(m!)^{2}}.
\end{eqnarray}
Hence
\begin{eqnarray}
|B^{(2)},\bk\rangle & = & \sqrt{\frac{(m!)^{2}}{(2m)!}}
\sum_{|\bn|=m}\sqrt{\frac{\bk !}{\bn !(\bk-\bn)!}}\,
|\bn\rangle\otimes |\bk-\bn\rangle.
\end{eqnarray}
The cases with larger $q$ can be treated in the same way.
The result is
\begin{eqnarray}
|B^{(q)}, \bk\rangle 
& = &
\sqrt{\frac{(m!)^{q}}{(mq)!}}\sum_{|\bn_{1}|=\dots=|\bn_{q}|=m}
\binom{\bk}{\bn_{1},\bn_{2},\dots,\bn_{q}}^{1/2}
|\bn_{1}\rangle\otimes |\bn_{2}\rangle\otimes\dots\otimes
|\bn_{q}\rangle,
\end{eqnarray}
where
\begin{eqnarray}
\binom{\bk}{\bn_{1},\bn_{2},\dots\bn_{q}} \equiv \left\{
\begin{array}{cc}
\frac{\bk !}{\bn_{1}!\,\bn_{2}!,\dots ,\bn_{q}!} &
(\mbox{if}\;\bn_{1}+\bn_{2}+\dots+\bn_{q}=\bk)\\
0 & (\mbox{else})
\end{array}\right. .
\end{eqnarray}

Suppose a state $|\varphi\rangle$ is represented in the occupation
number representation:
\begin{eqnarray}
|\varphi\rangle = \sum_{n}c_{\bk}\bigl|\bk\bigr\rangle.
\end{eqnarray}
From the general formulae (\ref{general1}) and (\ref{general2}), we obtain
\begin{eqnarray}
\tilde{M}^{(q)}_{|\varphi\rangle} = \sum_{|\bk|=mq}
\left|B_{\bk}\right|^{2},
\label{boson_general1}
\end{eqnarray}
\begin{eqnarray}
B_{\bk} = 
\sqrt{\frac{(m!)^{q}}{(mq)!}}\sum_{|\bn_{1}|=\dots=|\bn_{q}|=m}
\binom{\bk}{\bn_{1},\bn_{2},\dots,\bn_{q}}^{1/2}
c_{\bn_{1}}c_{\bn_{2}}\dots c_{\bn_{q}}. 
\label{boson_general2}
\end{eqnarray}

\section{Fermions}
\label{fermions}
\begin{figure}
\includegraphics[height=0.35\textwidth]{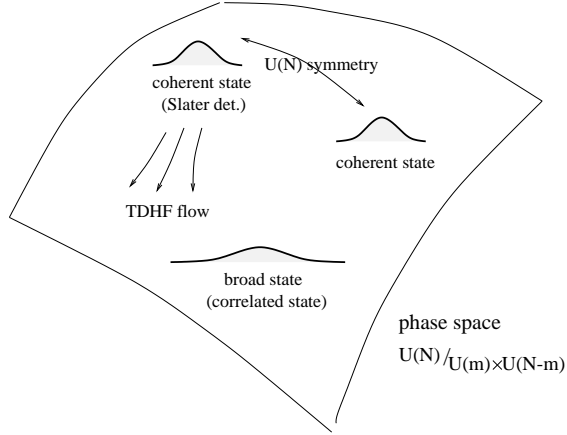}
\caption{Schematic picture of many-body Husimi distributions
in a fermionic system. 
The localized wave packets (coherent states) represent
independent-particle states (Slater determinants), 
and broad states represent
correlated states. All Slater determinants are connected by
the $U(N)$ group, and treated equally. The TDHF equation 
defines a flow in this phase space.}
\label{TDHF}
\end{figure}

In this section we study fermionic systems.

\subsection{Coherent states for fermionic systems}
We consider $m$ fermions in $N$ single-particle states.
The particle-hole excitation operators 
\begin{eqnarray}
X_{i}^{j} = a_{i}a_{j}^{\dagger}
\end{eqnarray}
satisfy the same equations (\ref{commut}) and (\ref{generate}) 
as in the bosonic case. 
Hence they generate the single-particle transformation
group $U(N)$. 
Fermionic many-particle states
form an irreducible representation which consists of anti-symmetric
combinations of the single-particle states. In the Young diagram, this 
representation is denoted by $[m]$ (See Fig. \ref{young_fermion}),
and we refer to it as $F[N,m]$.

As we have explained in the introduction,
the fermionic coherent states are Slater determinants,
which can be written as
\begin{eqnarray}
|F,\zeta\rangle = {\cal A}\left(|\varphi_{1}(\zeta)\rangle
|\varphi_{2}(\zeta)\rangle\dots |\varphi_{m}(\zeta)\rangle\right)
\end{eqnarray} 
where $|\varphi_{i}(\zeta)\rangle = U(\zeta)|\varphi_{i}\rangle$.
Unitary transformations in the subspaces spanned by
$|\varphi_{1}\rangle\dots |\varphi_{m}\rangle$
and $|\varphi_{m+1}\rangle\dots |\varphi_{N}\rangle$ are
irrelevant when we consider the coherent states.
Therefore the coherent states are specified
by a point on the complex Grassmann manifold
\begin{eqnarray}
U(N)/\left(U(N-m)\times U(m)\right)\simeq G_{m,N}(\mathbb{C}).
\end{eqnarray}
An explicit parametrization of the coherent states is
given by \cite{thouless}
\begin{eqnarray}
|F,\zeta\rangle = {\cal N}(\zeta)\exp
\left(\sum_{i=1}^{m}\sum_{j=m+1}^{N}\zeta_{j}^{i}
X_{i}^{j}\right)|0\rangle,
\end{eqnarray}
where $|0\rangle \equiv
|\overbrace{0,\dots,0}^{N-m},\overbrace{1,\dots,1}^{m}\rangle$ is
the fermionic ``vacuum'', and
$\zeta=\{\zeta_{j}^{i}\}$
is a coordinate system of $G_{m,N}(\mathbb{C})$.

The ``classical'' equation of motion defined on the manifold
is the TDHF (time-dependent Hartree-Fock) equation \cite{yamamura}.
(See Fig. \ref{TDHF}.) 
Therefore the manifold $G_{m,N}(\mathbb{C})$ 
is sometimes referred to as the TDHF manifold.
The Husimi distribution on this manifold is defined as
\begin{eqnarray}
{\cal H}_{|\varphi\rangle}(\zeta) = |\langle F,\zeta |\varphi\rangle|^{2}.
\end{eqnarray}

\subsection{Moments of the Husimi distribution}
\begin{figure}
\begin{picture}(200,150)(0,0)
\thicklines
\put(30,125){\framebox(15,15){}}
\put(30,110){\framebox(15,15){}}
\put(30,80){\framebox(15,30){\shortstack{$\cdot$\\$\cdot$\\$\cdot$}}}
\put(30,65){\framebox(15,15){}}
\put(120,125){\framebox(15,15){}}
\put(120,110){\framebox(15,15){}}
\put(120,80){\framebox(15,30){\shortstack{$\cdot$\\$\cdot$\\$\cdot$}}}
\put(120,65){\framebox(15,15){}}
\put(135,125){\framebox(40,15){$\cdot\cdot\cdot$}}
\put(135,110){\framebox(40,15){$\cdot\cdot\cdot$}}
\put(135,65){\framebox(40,15){$\cdot\cdot\cdot$}}
\put(175,125){\framebox(15,15){}}
\put(175,110){\framebox(15,15){}}
\put(175,80){\framebox(15,30){\shortstack{$\cdot$\\$\cdot$\\$\cdot$}}}
\put(175,65){\framebox(15,15){}}
\put(37.5,40){\makebox(0,0){$[m]$}}
\put(155,40){\makebox(0,0){$[m^{q}]=[\overbrace{m,m,\dots,m}^{q}]$}}
\put(37.5,10){\makebox(0,0){$F[N,m]$}}
\put(155,10){\makebox(0,0){$F^{(q)}[N,m]$}}
\end{picture}
\caption{Irreducible representations of $U(N)$,
$F[N,m]=F^{(1)}[N,m]$ and $F^{(q)}[N,m]$.}
\label{young_fermion}
\end{figure}
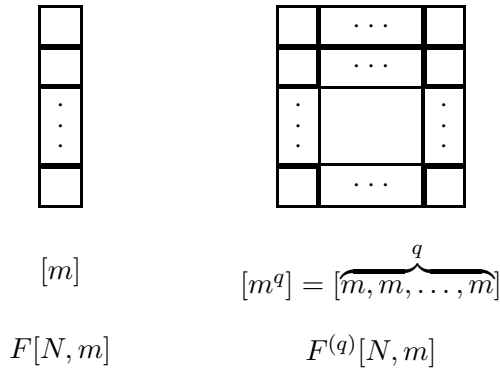

Let us consider a $m$-particle state
\begin{eqnarray}
|\varphi\rangle = \sum_{\bn}c_{\bn}|\bn\rangle,
\end{eqnarray}
where
\begin{eqnarray}
\bn = (n_{1},n_{2},\dots,n_{N})\;\;\; (n_{i} = 0\; {\rm or}\;1)
\end{eqnarray}
and
\begin{eqnarray}
|\bn| = \sum_{j}n_{j} = m.
\end{eqnarray}
We calculate the normalized $q$-th moment of the Husimi distribution
\begin{eqnarray}
\tilde{M}^{(q)}_{|\varphi\rangle} =
\frac{{\rm dim}\, D_{q\lambda}}{{\rm dim}\, D_{\lambda}}\,
M_{|\varphi\rangle}^{(q)}.
\end{eqnarray}
Here, $D_{q\lambda}$ is $[m^{q}]$ in
the Young diagram, which we refer to as $F^{(q)}[N,m]$ 
(Fig. \ref{young_fermion}).  In this case,
\begin{eqnarray}
{\rm dim}\,F^{(q)}=
\prod_{j=0}^{m-1}\frac{(N+q-j-1)!\,j!}{(N-j-1)!\,(q+j)!},
\end{eqnarray}
hence
\begin{eqnarray}
\frac{{\rm dim}\, D_{q\lambda}}{{\rm dim}\, D_{\lambda}}=
\prod_{j=0}^{m-1}\frac{\Gamma(N+q-j)\Gamma(j+2)}{\Gamma(N+1-j)\Gamma(q+j+1)}.
\end{eqnarray}

According to the general formulae (\ref{general1}) and (\ref{general2}),
\begin{eqnarray}
\tilde{M}^{(q)}_{|\varphi\rangle}=\sum_{\bk,\tau}
\left|B_{\bk,\tau}\right|^{2},
\end{eqnarray}
\begin{eqnarray}
B_{\bk,\tau} = \sum_{\bn}
\langle F^{(q)},\bk, \tau
|\left(|\bn_{1}\rangle\otimes\dots \otimes|\bn_{q}\rangle\right)
c_{\bn_{1}}\dots c_{\bn_{q}}.
\label{coeff_f}
\end{eqnarray}
Here an additional index $\tau$ is introduced because 
the occupation number $\bk$ does not necessarily specify a state in $F^{(q)}$.
The group-theoretical factors in (\ref{coeff_f}) can be evaluated,
for instance, by the eigenfunction method \cite{chen}.
However, it is not so easy as in the bosonic case
to obtain an explicit formula for them.
In the following, we calculate (\ref{coeff_f})
in some simple examples.

\subsection{Examples}

\begin{figure}
\begin{picture}(220,240)(0,-110)
\thicklines
\put (-5,95){\makebox(0,0){$F[3,1]$}}
\put (-15,55){\line(1,0){100}}
\put (35,5){\line(0,1){23}}
\put (35,37){\line(0,1){68}}
\put (85,60){\makebox(0,0){$H_{1}$}}
\put (42,105){\makebox(0,0){$H_{2}$}}
\put (55,66.5){\makebox(0,0){$|100\rangle$}}
\put (15,66.5){\makebox(0,0){$|010\rangle$}}
\put (35,31.9){\makebox(0,0){$|001\rangle$}}
\put (155,95){\makebox(0,0){$F[3,2]$}}
\put (145,55){\line(1,0){100}}
\put (195,5){\line(0,1){69}}
\put (195,83){\line(0,1){23}}
\put (245,60){\makebox(0,0){$H_{1}$}}
\put (202,105){\makebox(0,0){$H_{2}$}}
\put (215,43.5){\makebox(0,0){$|101\rangle$}}
\put (175,43.5){\makebox(0,0){$|011\rangle$}}
\put (195,78.1){\makebox(0,0){$|110\rangle$}}
\put (75,-15){\makebox(0,0){$F^{(2)}[3,2]$}}
\put (125,-105){\line(0,1){23}}
\put (125,-73){\line(0,1){60}}
\put (75,-55){\line(1,0){100}}
\put (175,-50){\makebox(0,0){$H_{1}$}}
\put (140,-15){\makebox(0,0){$H_{2}$}}
\put (165,-78.1){\makebox(0,0){$|202\rangle$}}
\put (125,-78.1){\makebox(0,0){$|112\rangle$}}
\put (85,-78.1){\makebox(0,0){$|022\rangle$}}
\put (145,-43.5){\makebox(0,0){$|211\rangle$}}
\put (105,-43.5){\makebox(0,0){$|121\rangle$}}
\put (125,-8.8){\makebox(0,0){$|220\rangle$}}
\end{picture}
\caption{Irreducible representations of $U(3)$. 
The weight vector $(H_{1}, H_{2})$ is related to occupation numbers by 
$H_{1} = (n_{1}-n_{2})/2$ and
$H_{2}=(n_{1}+n_{2}-2n_{3})/(2\sqrt{3})$. Single-particle
states  form the fundamental representation $F[3,1]=\bthree$,
and two-particle (i.e., one-hole) states form $F[3,2]=\bar{\bthree}$.
$F^{(2)}[3,2]$ corresponds to $\bar{\bsix}$.}
\label{rep_u3}
\end{figure}
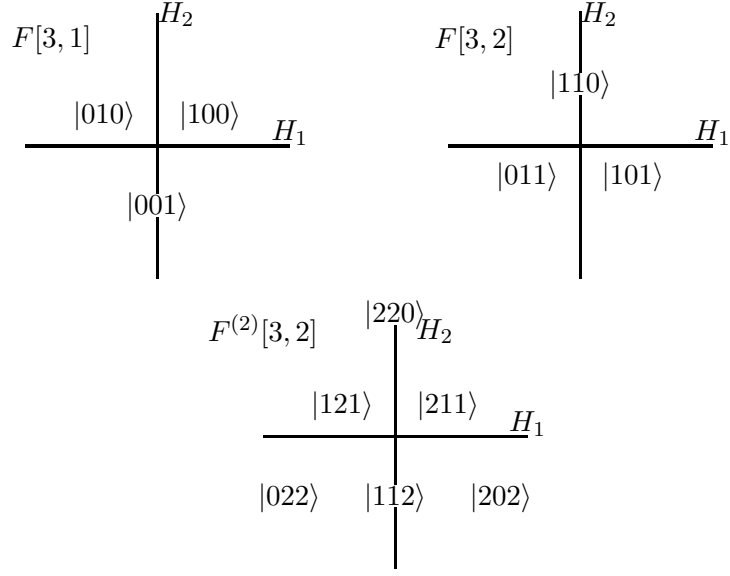

Let us consider some simple fermionic systems.
The cases with $m=1$ are trivial, where $\tilde{M}^{(2)}$
is always equal to unity. Let us consider the cases with $m=2$.
It is obviously trivial for $N=2$. For $N=3$, there are
three many-body states $|011\rangle$, $|101\rangle$
and $|110\rangle$, which form the representation
$\bar{\bthree}$ of $U(3)$ (See Fig. \ref{rep_u3}.)
The representation $F^{(2)}$ is $\bar{\bsix}$,
whose components are written as, for instance,
\begin{eqnarray}
|220\rangle & = & 
\left|\begin{array}{c}
110\\ 
110
\end{array}\right\rangle,\label{220}\\
|211\rangle & = & 
\frac{1}{\sqrt{2}}\left\{
\left|\begin{array}{c}
110\\ 
101
\end{array}\right\rangle
+
\left|\begin{array}{c}
101\\ 
110
\end{array}\right\rangle\right\}
\label{211}
\end{eqnarray}
Here we arranged the elements of the tensor products vertically.
For example, 
\begin{eqnarray}
\left|\begin{array}{c}
110\\ 
101
\end{array}\right\rangle
\equiv |110\rangle\otimes |101\rangle.
\end{eqnarray}
Note that the sum of the two row vectors gives 
the occupation number of the product.

All the components of $\bar{\bsix}$  
are obtained by permutating the
occupation numbers in (\ref{220}) and (\ref{211}). 
Then we obtain the components of $\tilde{M}^{(2)}$ as
\begin{eqnarray}
B_{220} & = & (c_{12})^{2},\\
B_{211} & = & \sqrt{2}\,c_{12}\,c_{13},\\
\dots.\nonumber
\end{eqnarray}
Here, $c_{ij}$ is the coefficient of the basis state where
$i$-th and $j$-th states are occupied. For example, $c_{12}$
is the coefficient of $|011\rangle$. 
The explicit formula for the second moment is
\begin{eqnarray}
\tilde{M}^{(2)} =
\left(|c_{12}|^{2} + |c_{23}|^{2} + |c_{31}|^{2}\right)^{2}=1.
\end{eqnarray}

This result is natural because $F[3,2]$ is a one-hole system,
and has no many-body correlation essentially.
By the same reason, $F[N,N-1]$ is also trivial for arbitrary $N$.
In this one-hole case, a state can be written as
\begin{eqnarray}
|\varphi\rangle = \sum_{j=1}^{N}c_{j}|j\rangle.
\end{eqnarray}
Here, $|j\rangle$ is a state with a hole in the $j$-th
single-particle state.
Since this system is essentially the
same as the one boson system $B[N,1]$, we can use 
the general formulae for bosons (\ref{boson_general2}). 
Then, by using (\ref{general3}) we obtain
\begin{eqnarray}
B_{\bk} & = & \sqrt{\frac{q!}{k_{1}!\,k_{2}!\dots k_{N}!}}
c_{1}^{k_{1}} c_{2}^{k_{2}}\dots c_{N}^{k_{N}}
\end{eqnarray}
and
\begin{eqnarray}
\tilde{M}^{(q)} = \left(\sum_{j}|c_{j}|^{2}\right)^{q} = 1.
\end{eqnarray}

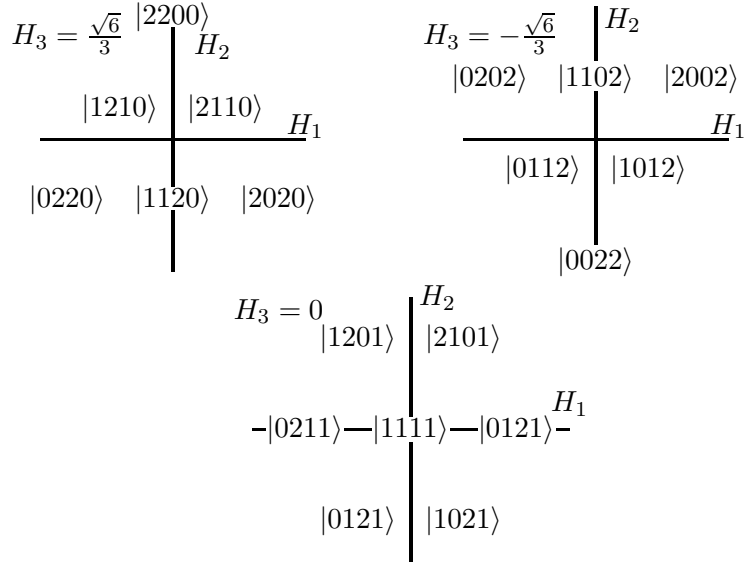
\begin{figure}
\begin{picture}(240,220)(0,-110)
\thicklines
\put (-5,95){\makebox(0,0){$H_{3}=\frac{\sqrt{6}}{3}$}}
\put (35,5){\line(0,1){23}}
\put (35,37){\line(0,1){60}}
\put (-15,55){\line(1,0){100}}
\put (85,60){\makebox(0,0){$H_{1}$}}
\put (50,90){\makebox(0,0){$H_{2}$}}
\put (75,31.9){\makebox(0,0){$|2020\rangle$}}
\put (35,31.9){\makebox(0,0){$|1120\rangle$}}
\put (-5,31.9){\makebox(0,0){$|0220\rangle$}}
\put (55,66.5){\makebox(0,0){$|2110\rangle$}}
\put (15,66.5){\makebox(0,0){$|1210\rangle$}}
\put (35,101.2){\makebox(0,0){$|2200\rangle$}}
\put (155,95){\makebox(0,0){$H_{3}=-\frac{\sqrt{6}}{3}$}}
\put (195,15){\line(0,1){58}}
\put (195,85){\line(0,1){20}}
\put (145,55){\line(1,0){100}}
\put (245,60){\makebox(0,0){$H_{1}$}}
\put (205,100){\makebox(0,0){$H_{2}$}}
\put (235,78.1){\makebox(0,0){$|2002\rangle$}}
\put (195,78.1){\makebox(0,0){$|1102\rangle$}}
\put (155,78.1){\makebox(0,0){$|0202\rangle$}}
\put (215,43.5){\makebox(0,0){$|1012\rangle$}}
\put (175,43.5){\makebox(0,0){$|0112\rangle$}}
\put (195,8.8){\makebox(0,0){$|0022\rangle$}}
\put (75,-10){\makebox(0,0){$H_{3}=0$}}
\put (125,-105){\line(0,1){45}}
\put (125,-50){\line(0,1){45}}
\put (65,-55){\line(1,0){5}}
\put (100,-55){\line(1,0){10}}
\put (140,-55){\line(1,0){10}}
\put (180,-55){\line(1,0){5}}
\put (185,-45){\makebox(0,0){$H_{1}$}}
\put (135,-5){\makebox(0,0){$H_{2}$}}
\put (145,-20.4){\makebox(0,0){$|2101\rangle$}}
\put (105,-20.4){\makebox(0,0){$|1201\rangle$}}
\put (145,-89.6){\makebox(0,0){$|1021\rangle$}}
\put (105,-89.6){\makebox(0,0){$|0121\rangle$}}
\put (165,-55){\makebox(0,0){$|0121\rangle$}}
\put (85,-55){\makebox(0,0){$|0211\rangle$}}
\put (125,-55){\makebox(0,0){$|1111\rangle$}}
\end{picture}
\caption{An irreducible representation of $U(4)$, $F^{(2)}[4,2]=\btwo\bzero$.
The weight vector $(H_{1}, H_{2},H_{3})$ is related to occupation numbers by 
$H_{1}=(n_{1}-n_{2})/2$,
$H_{2}=(n_{1}+n_{2}-2n_{3})/(2\sqrt{3})$ and
$H_{3}=(n_{1}+n_{2}+n_{3}-3n_{4})/(2\sqrt{6})$.
This representation forms a regular octahedron in the three-dimensional
weight space. 
$|1111\rangle$ is at the center thereof, and doubly degenerate.}
\label{rep_u4}
\end{figure}

The simplest non-trivial example is $F[4,2]$, whose dimension is
six. To calculate the second moment, we have to consider the CG series 
\begin{eqnarray}
\begin{array}{ccccccccc}
[2]&\otimes&[2]&=&[2,2]&\oplus&[3,1]&\oplus&[4]\\
\bsix & \otimes & \bsix & = & \btwo\bzero & \oplus 
& \bone\bfive & \oplus &  \bone .
\end{array}
\end{eqnarray}
The components of $F^{(2)}=[2,2]$ are written as, for instance, 
\begin{eqnarray}
|2200\rangle & = & 
\left|\begin{array}{c}
1100 \\
1100
\end{array}\right\rangle,\\
|2110\rangle & = & 
\frac{1}{\sqrt{2}}
\left\{\left|\begin{array}{c}
1010 \\
1100
\end{array}\right\rangle
+
\left|\begin{array}{c}
1010 \\
1100
\end{array}\right\rangle\right\}.
\end{eqnarray}
Note that $|1111\rangle$ is doubly degenerate.
(See Fig. \ref{rep_u4}.)
We define two basis vectors as
\begin{eqnarray}
|1111,a\rangle
& = & 
\frac{1}{2}\left\{
\left|\begin{array}{c}
1100 \\
0011
\end{array}\right\rangle
+
\left|\begin{array}{c}
0011 \\
1100
\end{array}\right\rangle
+
\left|\begin{array}{c}
1010 \\
0101
\end{array}\right\rangle
+
\left|\begin{array}{c}
0101 \\
1010
\end{array}\right\rangle\right\},\\
|1111,b\rangle & = & 
\frac{1}{2\sqrt{3}}\left\{
\left|\begin{array}{c}
1100 \\
0011
\end{array}\right\rangle
+
\left|\begin{array}{c}
0011 \\
1100
\end{array}\right\rangle
-
\left|\begin{array}{c}
1010 \\
0101
\end{array}\right\rangle
-
\left|\begin{array}{c}
0101 \\
1010
\end{array}\right\rangle\right.\nonumber\\
&& \left.
- 2
\left|\begin{array}{c}
1001 \\
0110
\end{array}\right\rangle
- 2
\left|\begin{array}{c}
0110 \\
1001
\end{array}\right\rangle\right\}.
\end{eqnarray}
Then
\begin{eqnarray}
B_{2200} & = & (c_{12})^{2},\\
B_{2110} & = & \sqrt{2}\,c_{12}\,c_{13},\\
B_{1111,a} & = & c_{12}c_{34} + c_{13}c_{24},\\
B_{1111,b} & = & \frac{c_{12}c_{34}
- c_{13}c_{24} - 2c_{14}c_{23}}{\sqrt{3}}.
\end{eqnarray}
The other components are obtained by permutating the occupation
numbers. Then we obtain
\begin{eqnarray}
\tilde{M}^{(2)} & = &
|c_{12}|^{4} + |c_{13}|^{4} + |c_{14}|^{4} +
|c_{23}|^{4} + |c_{24}|^{4} + |c_{34}|^{4} \nonumber\\
&& 
+ 2\left\{|c_{12}c_{13}|^{2} + |c_{12}c_{14}|^{2} + |c_{13}c_{14}|^{2}  
  + |c_{12}c_{23}|^{2} + |c_{12}c_{24}|^{2} + |c_{23}c_{24}|^{2}\right. 
\nonumber\\
&& \left.
  + |c_{13}c_{23}|^{2} + |c_{13}c_{34}|^{2} + |c_{23}c_{34}|^{2} 
  + |c_{14}c_{24}|^{2} + |c_{14}c_{34}|^{2} + |c_{24}c_{34}|^{2} 
\right\}\nonumber\\
&&
+ \frac{2}{3}\left\{
|c_{12}c_{34}+c_{13}c_{24}|^{2} + 
|c_{12}c_{34}-c_{14}c_{23}|^{2} +
|c_{13}c_{24}+c_{14}c_{23}|^{2}
\right\}
\label{m2_a}
\end{eqnarray}
A simpler expression of the second moment can be obtained from
(\ref{decompose}):
\begin{eqnarray}
\tilde{M}^{(2)} = 1 - \langle P_{[3,1]}\rangle 
- \langle P_{[4]}\rangle.
\end{eqnarray}
$\langle P_{[3,1]}\rangle$ vanishes identically because it 
is anti-symmetric with respect to the exchange of the 
components of the tensor product. Therefore all we have to
calculate is $\langle P_{[4]}\rangle$. The basis state of the
singlet representation
$[4]$ is
\begin{eqnarray}
|[4]\rangle = \frac{1}{\sqrt{6}}\left\{
\left|\begin{array}{c}
1100 \\
0011
\end{array}\right\rangle
+
\left|\begin{array}{c}
0011 \\
1100
\end{array}\right\rangle
-
\left|\begin{array}{c}
1010 \\
0101
\end{array}\right\rangle
-
\left|\begin{array}{c}
0101 \\
1010
\end{array}\right\rangle
+
\left|\begin{array}{c}
1001 \\
0110
\end{array}\right\rangle
+
\left|\begin{array}{c}
0110 \\
1001
\end{array}\right\rangle
\right\}.
\end{eqnarray}
Hence 
\begin{eqnarray}
\tilde{M}^{(2)} & = & 1-\langle P_{[4]}\rangle = 
1 - \frac{2}{3}\left|c_{12}c_{34}-c_{13}c_{24}+c_{14}c_{23}\right|^{2}. 
\end{eqnarray}

The second moment of the general two-fermion case $F[N,2]$ ($N\ge 4$)
is
obtained by a similar calculation.
The result is   
\begin{eqnarray}
\tilde{M}^{(2)} & = & 
 \sum_{i_{1},i_{2}}|c_{i_{1},i_{2}}|^{4}
+ 2\sum_{j_{1},j_{2},k}|c_{k,j_{1}}c_{k,j_{2}}|^{2} 
 + \sum_{l_{1},l_{2},l_{3},l_{4}}A(l_{1},l_{2},l_{3},l_{4}).
\end{eqnarray}
Here,
\begin{eqnarray}
A(i,j,k,l)&\equiv& 
\frac{2}{3}\left\{
|c_{ij}c_{kl}+c_{ik}c_{jl}|^{2} + 
|c_{ij}c_{kl}-c_{il}c_{jk}|^{2} +
|c_{ik}c_{jl}+c_{il}c_{jk}|^{2}
\right\},
\end{eqnarray} 
and the sum is taken over $1\le i_{1}<i_{2}\le N$, 
$1\le j_{1}<j_{2}\le N$, $1\le k\le N$ and
$1\le l_{1}<l_{2}<l_{3}<l_{4}\le N$. 
This can be rewritten as 
\begin{eqnarray}
\tilde{M}^{(2)} & = & 1 - \frac{2}{3}
\sum_{l_{1},l_{2},l_{3},l_{4}}
\left|c_{l_{1}l_{2}}c_{l_{3}l_{4}}-c_{l_{1}l_{3}}
c_{l_{2}l_{4}}+c_{l_{1}l_{4}}c_{l_{2}l_{3}}\right|^{2} 
\end{eqnarray}
by using the constraint $\sum_{i_{1}i_{2}}|c_{i_{1}i_{2}}|^{2}=1$.

\section{Distinguishable particles}
\label{distinguishable}

\subsection{General formalism}
Although all elementary particles are considered to be bosons or 
fermions, we often treat particles as distinguishable if the
exchange effect is negligible. 
When we treat distinguishable particles, we can change the
single-particle basis of each particle separately. 
If there are
$m$ particles and the $i$-th particle can take 
$N_{i}$ states, 
the group of the single-particle transformation is
\begin{eqnarray}
SU(N_{1})\times SU(N_{2})\times\cdots\times SU(N_{m}).
\end{eqnarray}
We denote the Hilbert space of this system 
as $D[N_{1},N_{2},\dots,N_{m}]$, whose dimension is
$\prod_{i=1}^{m}N_{i}$. It can also be written as
$\overbrace{[1]\otimes [1]\otimes \dots\otimes [1]}^{m}$, where 
$i$-th $[1]$ is the fundamental representation of $SU(N_{i})$.
A coherent states of this system is a separable 
(disentangled) state, which is considered to be a point on
the ``phase space'' $\mathbb{C}P^{N_{1}-1}\times \mathbb{C}P^{N_{2}-1}
\times\dots\times\mathbb{C}P^{N_{m}-1}$. 
The Husimi distribution is defined as a distribution function
on this phase space.

\subsection{Moments}

The normalized $q$-th moment is defined as
\begin{eqnarray}
\tilde{M}^{(q)}_{|\varphi\rangle} = 
\frac{{\rm dim}\, D_{q\lambda}}{{\rm dim}\, D_{\lambda}}
M^{(q)}_{|\varphi\rangle}.
\end{eqnarray}
Here, 
$D_{q\lambda}=[1^{q}]\otimes\dots\otimes [1^{q}]$,
which we denote as $D^{(q)}[N_{1},\dots,N_{m}]$.
We will drop $[N_{1},\dots,N_{m}]$ when it is obvious.

Let us calculate the normalized moment by using the general
formulae (\ref{general1}) and (\ref{general2}).
If $m=1$, this is the same as the bosonic case with $m=1$.
A state in $[1^{q}]$ is represented by occupation numbers
$\bk=(k_{1},\dots,k_{N_{1}})$, which satisfies $|\bk|=q$.
Then 
\begin{eqnarray}
\langle D^{(q)},\bk |
\bigl(|\bn_{1}\rangle\otimes\dots\otimes |\bn_{q}\rangle\bigr)
= \frac{1}{\sqrt{q!}}
\binom{\bk}{\bn_{1},\dots,\bn_{N_{1}}}^{\frac{1}{2}}.
\label{m=1p}
\end{eqnarray}
Here, $|D^{(q)},\bk\rangle \in [1^{q}]$ and $|\bn_{i}\rangle \in [1]$.
Since $\bn_{i}!=1$, (\ref{m=1p}) can be written simply as
\begin{eqnarray}
\langle D^{(q)},\bk|
\bigl(|\bn_{1}\rangle\otimes\dots\otimes|\bn_{q}\rangle\bigr)
= \left\{\begin{array}{cc}
\sqrt{\frac{\bk!}{q!}} & 
\left(\mbox{if}\,\sum_{l=1}^{N_{1}}\bn_{l}=\bk\right)\\
0 & (\mbox{else})
\end{array}\right. .
\label{m=1}
\end{eqnarray}

The coefficients in (\ref{general2}) for multi-particle cases
are easily obtained from (\ref{m=1}).
A basis state in $D^{(q)}[N_{1},\dots,N_{m}]$ is specified by
a set of occupation numbers $\{k_{i,j_{i}}\}$, where
$1\le i\le m$, $1\le j_{i}\le N_{i}$ and $\sum_{j_{i}}k_{i,j_{i}} = q$
for $\forall i$. Then we have
\begin{eqnarray}
\langle D^{(q)},\{k_{i,j_{i}}\}|
\bigl(\left|\{n_{i,j_{i},1}\}\right\rangle\otimes\dots\otimes
|\{n_{i,j_{i},q}\}\rangle\bigr)
= \left\{\begin{array}{cc}
\sqrt{\frac{\prod_{i,j_{i}}k_{i,j_{i}}!}{(q!)^{m}}} & 
\left(\mbox{if}\,\sum_{l=1}^{N_{i}}n_{i,j_{i},l}=k_{i,j_{i}}\,
\mbox{for}\, \forall (i,j_{i})\right)\\
0 & (\mbox{else})
\end{array}\right. ,
\end{eqnarray}
where $|D^{(q)},\{k_{i,j_{i}}\}\rangle\in D^{(q)}[N_{1},\dots,N_{m}]$
and $|\{n_{i,j_{i},l}\}\rangle \in D[N_{1},\dots,N_{m}]$.

\subsection{Qubits}
In this and the next 
subsections, we investigate systems of qubits rather in detail.
Let us consider a system of $m$-qubits, which is
$D[\overbrace{2,2,\dots,2}^{m}]$ in our notation.
This is the spin $(1/2,1/2,\dots,1/2)$ representation of 
$SU(2)\times SU(2)\times\dots\times SU(2)$, and 
$D^{(q)}[2,2,\dots,2]$ is the spin $(q/2,q/2,\dots,q/2)$ representation 
thereof. 
We denote a basis state of this system
as $|\bmu\rangle\equiv|\mu_{1},\mu_{2},\dots,\mu_{m}\rangle$, 
where $\mu_{i}$ takes
$1/2$ or $-1/2$. This notation is better than using 
$1$ and $0$ in that the symmetry between the two states
is obvious. In the same way, a basis state in 
$D^{(q)}[2,2,\dots,2]$ is denoted as 
$|\mu_{1},\mu_{2},\dots,\mu_{m}\rangle$, where 
$\mu_{i}$ runs from $-q/2$ to $q/2$ with the unity steps. 
This is related to the occupation number representation in the 
previous subsection by
\begin{eqnarray}
n_{i,1} & = & \frac{q}{2}+\mu_{i},\\
n_{i,2} & = & \frac{q}{2}-\mu_{i}.
\end{eqnarray}

According to the results in the previous subsection,
the normalized $q$-th moment
\begin{eqnarray}
\tilde{M}^{(q)}_{|\varphi\rangle} = 
\left(\frac{q+1}{2}\right)^{m}
M^{(q)}_{|\varphi\rangle}
\end{eqnarray}
is obtained as
\begin{eqnarray}
\tilde{M}^{(q)}_{|\varphi\rangle} =
\sum_{\bnu}|B_{\bnu}|^{2},
\end{eqnarray}
\begin{eqnarray}
B_{\bnu} = \sum_{\bn_{1},\dots,\bn_{q}}
\langle D^{(q)},\bnu |\bigl(|\bmu_{1}\rangle\otimes 
|\bmu_{2}\rangle\otimes\dots\otimes |\bmu_{q}\rangle\bigr)
c_{\bmu_{1}}\dots c_{\bmu_{q}}
\end{eqnarray}
and
\begin{eqnarray}
\langle D^{(q)},\bnu |\bigl(|\bmu_{1}\rangle\otimes 
|\bmu_{2}\rangle\otimes\dots\otimes |\bmu_{q}\rangle\bigr)
= \left\{\begin{array}{cc}
\prod_{i=1}^{m}\binom{q}{\frac{q}{2}+\nu_{i}}^{-1/2} & 
\left(\mbox{If}\, \sum_{j=1}^{q}\bmu_{j}=\bnu\right)\\
0 & (\mbox{else})
\end{array}\right. .
\label{general_qubit}
\end{eqnarray}

\subsection{Examples}

Let us start with a single qubit case $D[2]$. 
A qubit state is represented as
\begin{eqnarray}
|\varphi\rangle = c_{+}|+\rangle + c_{-}|-\rangle,
\end{eqnarray}
where $+$ and $-$ are abbreviations of $+1/2$ and $-1/2$.
The normalized $q$-th moment of the state is
\begin{eqnarray}
\tilde{M}^{(q)}_{|\varphi\rangle} &=& \sum_{\nu=-q/2}^{q/2}|B_{\nu}|^{2},
\end{eqnarray}
where $B_{\nu}$ is obtained from 
(\ref{general3}) and (\ref{general_qubit}) as
\begin{eqnarray}
B_{\nu} = 
\binom{q}{\frac{q}{2}+\nu}^{1/2}{c_{+}}^{q/2+\nu}{c_{-}}^{q/2-\nu}.
\end{eqnarray}
Therefore
\begin{eqnarray}
\tilde{M}^{(q)} 
& = & \left(|c_{+}|^{2}+|c_{-}|^{2}\right)^{q} = 1.
\end{eqnarray}
This case is trivial, as it should be.

Next we consider 2-qubit case $D[2,2]$.
A state in this space is represented as
\begin{eqnarray}
|\varphi\rangle = c_{++}|++\rangle
+c_{+-}|+-\rangle + c_{-+}|-+\rangle + c_{--}|--\rangle.
\end{eqnarray}
$D^{(2)}[2,2]$ is a 9-dimensional representation, whose basis states are
specified by $\bnu =(\nu_{1},\nu_{2})$ $(-1\le\nu_{i}\le 1)$.
The coefficients $B_{\bnu}$  are easily
obtained as
\begin{eqnarray}
B_{11} & = & (c_{++})^{2},\\
B_{10} & = & \sqrt{2}\,c_{++}c_{+-},\\
B_{00} & = & c_{++}c_{--} + c_{+-}c_{-+},\\
&&\dots \nonumber
\end{eqnarray}
The other components which are not explicitly written here
are obtained by exchanging the first and the second suffixes and
changing their signs 
($+\leftrightarrow -$, $1\leftrightarrow -1$).

The explicit form of the normalized second moment is
\begin{eqnarray}
\tilde{M}^{(2)}_{|\varphi\rangle} & = & 
|c_{++}|^{4}+|c_{+-}|^{4}+|c_{-+}|^{4}+|c_{--}|^{4}\nonumber\\
&& 
+ 2|c_{++}c_{+-}|^{2} + 2|c_{++}c_{-+}|^{2}
+ 2|c_{--}c_{+-}|^{2} + 2|c_{--}c_{-+}|^{2}\nonumber\\
&&
+ |c_{++}c_{--}+c_{+-}c_{-+}|^{2}.
\end{eqnarray}
Let us rewrite this by using (\ref{decompose}) as
\begin{eqnarray}
\tilde{M}^{(2)}_{|\varphi\rangle} & = & 
1 - \langle P_{(1,0)}\rangle - \langle P_{(0,1)}\rangle
- \langle P_{(0,0)}\rangle,
\label{rewritten}
\end{eqnarray} 
where $(J_{1},J_{2})$ represents the irreducible 
representation with the total spins $J_{1}$ and $J_{2}$.
The spin 1 representation, which has three components 
\begin{eqnarray}
|1\rangle = 
\left|\begin{array}{c}
+\\+
\end{array}\right\rangle,\;\;\;
|0\rangle =  \frac{1}{\sqrt{2}}\left\{
\left|\begin{array}{c}
+\\-
\end{array}\right\rangle
+
\left|\begin{array}{c}
-\\+
\end{array}\right\rangle\right\},\;\;\;
|-1\rangle = 
\left|\begin{array}{c}
-\\-
\end{array}\right\rangle,
\end{eqnarray}
are symmetric with respect to the exchange between the components
of the tensor product. The spin 0 representation, whose basis state
is
\begin{eqnarray}
|0\rangle =  \frac{1}{\sqrt{2}}\left\{
\left|\begin{array}{c}
+\\-
\end{array}\right\rangle
-
\left|\begin{array}{c}
-\\+
\end{array}\right\rangle\right\},
\end{eqnarray}
is anti-symmetric with respect to the exchange.
Note that we arranged the components of the tensor product
vertically so that we can distinguish it from the tensor
product to represent the physical composite systems.
$\langle P_{(1,0)}\rangle$ and $\langle P_{(0,1)}\rangle$ vanish
because the representations $(1,0)$ and $(0,1)$ are anti-symmetric as a whole,
and only $\langle P_{(0,0)}\rangle$ survives in (\ref{rewritten}).
The basis state of the singlet representation $(0,0)$ is
\begin{eqnarray}
|00\rangle &=& \frac{1}{\sqrt{2}}\left\{
\left|\begin{array}{c}
+\\-
\end{array}\right\rangle
-
\left|\begin{array}{c}
-\\+
\end{array}\right\rangle
\right\} \otimes \frac{1}{\sqrt{2}}\left\{
\left|\begin{array}{c}
+\\-
\end{array}\right\rangle
-
\left|\begin{array}{c}
-\\+
\end{array}\right\rangle
\right\} \\
&=&
\frac{1}{2}\left\{
\left|\begin{array}{c}
++\\--
\end{array}\right\rangle
+
\left|\begin{array}{c}
--\\++
\end{array}\right\rangle
-
\left|\begin{array}{c}
+-\\-+
\end{array}\right\rangle
-
\left|\begin{array}{c}
-+\\+-
\end{array}\right\rangle
\right\}.
\label{(0,0)}
\end{eqnarray}
Then we obtain
\begin{eqnarray}
\langle P_{(0,0)}\rangle = |c_{++}c_{--}-c_{+-}c_{-+}|^{2}
= \frac{C^{2}}{4},
\end{eqnarray}
where $C$ is the concurrence \cite{hill}. Hence
\begin{eqnarray}
\tilde{M}^{(2)}_{|\varphi\rangle} = 1 - \frac{C^{2}}{4}.
\end{eqnarray}

The next example is the second moment of the $3$-qubit case.
\begin{eqnarray}
B_{111} & = & (c_{+++})^{2}, \\
B_{110} & = & \sqrt{2}c_{+++}c_{++-}, \\
B_{100} & = & c_{+++}c_{+--} + c_{++-}c_{+-+}, \\
B_{000} & = & \frac{1}{\sqrt{2}}
(c_{+++}c_{---} + c_{++-}c_{--+}
+ c_{+-+}c_{-+-} + c_{-++}c_{+-}).
\end{eqnarray}
The other components are obtained by permutating the suffixes and
changing their signs. Then we obtain
\begin{eqnarray}
\tilde{M}^{(2)}_{|\varphi\rangle} & = & 
|c_{+++}|^{4} + |c_{---}|^{4} + |c_{++-}|^{4} + |c_{+-+}|^{4} + 
|c_{-++}|^{4} + |c_{--+}|^{4} + |c_{-+-}|^{4} + |c_{+--}|^{4}
\nonumber\\
&& +
2|c_{+++}c_{++-}|^{2} + 2|c_{+++}c_{+-+}|^{2} +
2|c_{+++}c_{-++}|^{2} + 2|c_{---}c_{--+}|^{2} 
\nonumber\\
&& +
2|c_{---}c_{-+-}|^{2} + 2|c_{---}c_{+--}|^{2} +
2|c_{+-+}c_{+--}|^{2} + 2|c_{-++}c_{--+}|^{2} 
\nonumber\\
&& +
2|c_{++-}c_{-+-}|^{2} + 2|c_{-++}c_{-+-}|^{2} 
+ 2|c_{++-}c_{+--}|^{2} + 2|c_{+-+}c_{--+}|^{2}
\nonumber\\
&& +
|c_{+++}c_{+--}+c_{++-}c_{+-+}|^{2} +
|c_{+++}c_{--+}+c_{+-+}c_{-++}|^{2} +
|c_{+++}c_{-+-}+c_{-++}c_{++-}|^{2} 
\nonumber\\
&& +
|c_{---}c_{-++}+c_{--+}c_{-+-}|^{2} +
|c_{---}c_{++-}+c_{-+-}c_{+--}|^{2} +
|c_{---}c_{+-+}+c_{+--}c_{--+}|^{2} 
\nonumber\\
&& +
\frac{1}{2}|c_{+++}c_{---}+c_{++-}c_{--+}+c_{+-+}c_{-+-}+c_{-++}c_{+--}|^{2}.
\end{eqnarray}

By using (\ref{decompose}), $\tilde{M}^{(2)}$ can also be written as
\begin{eqnarray}
\tilde{M}^{(2)} = 1 
- \langle P_{(0,0,1)}\rangle
- \langle P_{(0,1,0)}\rangle
- \langle P_{(1,0,0)}\rangle.
\end{eqnarray}
Here we dropped the anti-symmetric representations $(1,1,0)$,
$(1,0,1)$, $(0,1,1)$ and $(0,0,0)$.
$(0,0,1)$ is a 3-dimensional representation, whose basis states are
\begin{eqnarray}
|001\rangle & = &
\frac{1}{2}\left\{
\left|\begin{array}{c}
+++ \\ --+
\end{array}\right\rangle
+
\left|\begin{array}{c}
--+ \\ +++
\end{array}\right\rangle
-
\left|\begin{array}{c}
+-+ \\ -++
\end{array}\right\rangle
-
\left|\begin{array}{c}
-++ \\ +-+
\end{array}\right\rangle\right\},\\
|000\rangle & = &
\frac{1}{2\sqrt{2}}\left\{
\left|\begin{array}{c}
+++ \\ ---
\end{array}\right\rangle
+
\left|\begin{array}{c}
--+ \\ ++-
\end{array}\right\rangle
-
\left|\begin{array}{c}
+-+ \\ -+-
\end{array}\right\rangle
-
\left|\begin{array}{c}
-++ \\ +--
\end{array}\right\rangle\right.\nonumber\\
&& \left.
+
\left|\begin{array}{c}
++- \\ --+
\end{array}\right\rangle
+
\left|\begin{array}{c}
--- \\ +++
\end{array}\right\rangle
-
\left|\begin{array}{c}
+-- \\ -++
\end{array}\right\rangle
-
\left|\begin{array}{c}
-+- \\ +-+
\end{array}\right\rangle
\right\},\\
|00\! -\!1\rangle & = &
\frac{1}{2}\left\{
\left|\begin{array}{c}
++- \\ ---
\end{array}\right\rangle
+
\left|\begin{array}{c}
--- \\ ++-
\end{array}\right\rangle
-
\left|\begin{array}{c}
+-- \\ -+-
\end{array}\right\rangle
-
\left|\begin{array}{c}
-+- \\ +--
\end{array}\right\rangle\right\}.
\end{eqnarray}  
Hence
\begin{eqnarray}
\langle P_{(0,0,1)}\rangle &=& \frac{1}{2}
|c_{+++}c_{--+}-c_{+-+}c_{-++}|^{2} + |c_{++-}c_{---}-c_{+--}c_{-+-}|^{2}\\
&&
+ \frac{1}{2}|c_{+++}c_{---} + c_{++-}c_{--+}
- c_{+-+}c_{-+-} - c_{+--}c_{-++}|^{2}. 
\end{eqnarray}
By explicit calculation this is shown to be equal to 
$\frac{1}{4}{\rm Tr}\left(\rho_{AB}\tilde{\rho}_{AB}\right)$.
Here, A,B and C are
the three qubits. 
$\rho_{AB}$ is the density matrix of the
pair A and B which is obtained by tracing out the information of 
C. $\tilde{\rho}_{AB}$ is the ``spin-flipped'' density matrix 
\cite{wootters}
\begin{eqnarray}
\tilde{\rho}_{AB} = (\sigma_{y}\otimes\sigma_{y})
\rho_{AB}^{*}(\sigma_{y}\otimes\sigma_{y}),
\end{eqnarray}
where the asterisk denotes complex conjugation in the standard basis
and 
\begin{eqnarray}
\sigma_{y} = \left(\begin{array}{cc}
0 & -i\\
i & 0
\end{array}\right).
\end{eqnarray}

According to \cite{coffman}, 
\begin{eqnarray}
{\rm Tr}\left(\rho_{AB}\tilde{\rho}_{AB}\right) = C_{AB}^{2}
+ \frac{1}{2}\tau_{ABC},
\end{eqnarray}
where $C_{AB}$ is the concurrence of the density matrix
$\rho_{AB}$ and $\tau_{ABC}$ is the 3-tangle whose explicit form
is given in \cite{coffman}. In the same way we can calculate
$\langle P_{(0,1,0)}\rangle$ and $\langle P_{(0,0,1)}\rangle$,
and obtain the result
\begin{eqnarray}
\tilde{M}^{(2)} = 1 - 
\frac{1}{4}(C_{AB}^{2}+C_{BC}^{2}+C_{CA}^{2})
- \frac{3}{8}\tau_{ABC}.
\end{eqnarray}

\section{Concluding remarks}
\label{conclude}
In this paper, we have defined generalized Husimi distributions
for many-body systems, and showed that their moments can be 
practically useful measures of complexity of many-body 
quantum states. We have derived an algebraic formula
to evaluate the moments with integer indices.
For bosons and distinguishable particles we have
derived the group theoretical factors appearing in
the formula explicitly.
We have also examined some special cases of fermions,
though we have not yet obtained a general formula for the 
group-theoretical factors in fermionic cases. 

The moments take the maximum values if and
only if the state is an independent-particle state. 
Therefore the moments are considered to be measures
of entanglement.
Although a lot of measures of entanglement have been
proposed so far, our method has an advantage of 
being able to produce infinitely many measures systematically.
Of course, not all of them are independent.
Nevertheless, it might be possible that the moments
and the related invariants (See (\ref{decompose}))
form a complete set of algebraic invariants
to classify the pure state entanglement.

The relation between delocalization of the Husimi distribution
and chaoticity of the classical mechanics has been shown
in one-body systems \cite{sugita, takahashi}.
We can expect a similar correspondence also in many-body 
systems, which means a relation between many-body
correlation and chaoticity of the dynamics of coherent
states (mean fields). 
For example, chaotic behaviors in fermionic systems like nuclei have been 
studied from the viewpoints of the shell model \cite{zelevinsky}
and the mean field dynamics
\cite{yamamura}.
However, the relation between the two standpoints has been unclear.
We hope that the Husimi distribution will be a bridge between
the two fields.

In this paper, we have concerned only with the systems
with definite particle numbers, and considered
classifications based on the single-particle unitary transformations. 
When we treat a systems with variable particle number,
we can introduce a generalized one-body transformation, 
which is known as the Bogoliubov transformation.
Then, for example, BCS-type wavefunctions are considered
to be coherent states in fermionic systems.
The treatment of such systems will be reported elsewhere.

\section*{Acknowledgment}
The author thanks K. \.{Z}yczkowski for helpful 
suggestions and discussions.
He also thanks Y. Fujiwara, A. Hosaka and 
V. Dmitra\v{s}inovi\'{c} for their comments about
group theory.


\begin{thebibliography}{99}

\bibitem{sugita} A. Sugita and H. Aiba, 
Phys. Rev. E. {\bf 65} (2002) 036205.

\bibitem{husimi}
K. Husimi, Proc. Phys. Math. Soc. Japan {\bf 22} 264 (1940).

\bibitem{perelomov}A. Perelomov, {\it Generalized Coherent States and
Their Applications}, Springer-Verlag Berlin Heidelberg, (1986).

\bibitem{zyczkowski} S. Gnutzmann and K. \.{Z}yczkowski,
J. Phys. A {\bf 34} (2001) 10123.

\bibitem{wehrl}
A. Wehrl, Rep. Math. Phys. {\bf 16} (1979) 353.

\bibitem{lieb}
E. H. Lieb, Commun. Math. Phys. {\bf 62} (1978) 35.

\bibitem{sugita2}
A. Sugita,
J. Phys. A: Math. Gen. {\bf 35} (2002) L621.

\bibitem{zhang}
W. M. Zhang and D. H. Feng, Phys. Rep. {\bf 252} (1995), 1. 

\bibitem{thouless}
D. J. Thouless, Nucl. Phys. {\bf 21} (1960), 225. 

\bibitem{yamamura}
M. Yamamura and A. Kuriyama, Prog. Theor. Phys Suppl. {\bf 93} (1987).

\bibitem{chen}
Jin-Quan Chen, {\it Group Representation Theory for Physicists},
World Scientific, (1989).

\bibitem{hill}
S. Hill and W. K. Wootters, 
Phys. Rev. Lett. {\bf 78} (1997) 5022.

\bibitem{wootters}
W. K. Wootters,
Phys. Rev. Lett. {\bf 80} (1998) 2245.

\bibitem{coffman}
V. Coffman, J. Kundu and W. K. Wootters,
Phys. Rev. A {\bf 61} (2000) 052306.

\bibitem{takahashi}
K. Takahashi, Suppl. Prog. Theor. Phys. {\bf 98} (1989) 109.

\bibitem{zelevinsky}
V. Zelevinsky, B. A. Brown, N. Frazier and M. Horoi,
Phys. Rep. {\bf 276} (1996) 85.

\end{thebibliography}
\end{document}